\documentclass[aps,prd,reprint,superscriptaddress,floatfix,twocolumn,nofootinbib]{revtex4-2}

\usepackage{color}
\usepackage{amssymb}
\usepackage{graphicx,color}
\usepackage[usenames,dvipsnames]{xcolor}
\usepackage[section]{placeins}
\usepackage{amsmath}
\usepackage{subfigure}
\usepackage[normalem]{ulem}
\usepackage{booktabs}
\usepackage{xcolor}
\usepackage{epsfig,graphics,color,graphicx,amsmath}
\usepackage{enumitem}

\colorlet{darkgreen}{green!50!black}
\colorlet{brightyellow}{yellow!75!red}
\colorlet{orange}{red!50!yellow}
\colorlet{darkblue}{blue!60!black}
\colorlet{darkred}{red!80!black}

\newcommand{\pslash}{\slash \hspace{-0.22cm} p}
\newcommand{\kslash}{\slash \hspace{-0.22cm} k}

\newcommand{\ppslash}{\slash \hspace{-0.224cm} p}

\usepackage{soul}

\usepackage{mathtools}
\usepackage{mathrsfs}
\usepackage{bm}
\usepackage{relsize}
\usepackage{indentfirst}

\usepackage{xcolor}

\usepackage{amssymb,amsmath,multirow,epsfig,graphicx,color}

\usepackage{epsfig}
\usepackage{graphicx,amsmath}
\def\be{\begin{eqnarray} &&}

\def\ee{\end{eqnarray}}
\def\psla{\slash \! \!\! }

\begin{document}

\title{ Pion inspired by QCD: Nakanishi and  Light-Front Integral Representations  }
\author{R.~M.~Moita}
\affiliation{Instituto Tecnol\'ogico de Aeron\'autica,  DCTA,
12228-900 S\~ao Jos\'e dos Campos,~Brazil}
\author{J.~P.~B.~C.~de Melo}
\affiliation{Laborat\'orio de F\'\i sica Te\'orica e 
Computacional - LFTC, 
\\
Universidade Cruzeiro do Sul and 
Universidade Cidade de S\~ao Paulo (UNICID) 
\\  01506-000 S\~ao Paulo, Brazil}
\author{T.~Frederico}
\affiliation{Instituto Tecnol\'ogico de Aeron\'autica,  DCTA,
12228-900 S\~ao Jos\'e dos Campos,~Brazil}
\author{W.~de Paula}
\affiliation{Instituto Tecnol\'ogico de Aeron\'autica,  DCTA,
12228-900 S\~ao Jos\'e dos Campos,~Brazil}

\begin{abstract}
 The pion structure in Minkowski space is explored using the Nakanishi integral representation. A general framework is developed for  the  pion Bethe-Salpeter amplitude based on the K\"allen-Lehmann representation of the dressed quarks with an ansatz for the pseudo-scalar $\pi-q\bar q$ vertex fulfilling the axial Ward-Takahashi identity. The Nakanishi weight functions are derived for the scalar amplitudes associated with the decomposition of the pion Bethe-Salpeter amplitude in different operator bases in the Dirac spinor space in terms of the involved spectral densities.
 The approach is applied to  an analytical model of the pion Bethe-Salpeter amplitude, which is
 combined with Landau gauge lattice  QCD results for the quark running mass at space-like momentum.   From the Nakanishi integral representation several  pion observables  were calculated, such as the decay constant, the spin decomposition of  the valence probabilities, longitudinal and transverse momentum distributions from the valence component of the light-front wave function. 
\end{abstract}

\date{\today}
\maketitle

\section{Introduction}

The intricate infrared (IR) and non-perturbative physics of Quantum Chromodynamics (QCD), which is expected to describe the hadron phenomenology from low to high energies, is under intense research both theoretically and experimentally. Understanding the richness of the strong interaction, reflected in the possibility of 3D imaging of hadrons  in terms of their quark and gluon content, is  among the motivations to built new electron-ion collider  (see e.g.~\cite{khalek2021science}).  

The  IR physics of QCD evolves the hadron partonic description in terms of the fundamental degrees of freedom of quarks and gluons to dressed constituents. While  heavy  quarks acquires the bulk of their masses  from the coupling to the Higgs field~\cite{Higgs:1964pj},  this mechanism is not enough  to give mass to the light flavors and does not explain the nucleon mass. On the other hand, the  strong IR interaction, as shown by  lattice QCD (LQCD calculations dresses both light quarks~\cite{ParPRD06} and  gluons~\cite{OLJPG11}),  as also found by continuum Euclidean methods using truncated Dyson-Schwinger~\cite{CloPPNP14,Eichmann:2016yit,Roberts:2021nhw}. As a result, at low momentum scales the light quarks become massive, and consequently, the  hadrons composed by them gain their weight. This mechanism corresponds to the dynamical chiral symmetry breaking   (DCSB), where the pion and kaon are identified as the associated Goldstone bosons~\cite{horn2016pion,Aguilar:2019teb}.

The infrared non-perturbative dynamics of QCD  at long distances, responsible for the color confinement, is also understood as the origin of DCSB with the light quarks dressed by the strong interaction. They are understood as  effective degrees of freedom associated with constituent quarks, which are widely used in the literature for modeling both the spectrum and the structure of hadrons (see e.g. the review of the quark model in~\cite{Zyla:2020zbs}).
In particular, the constituent quark model, formulated on the light-front (LF), is a practical theoretical approach to study non-perturbative problems in hadron structure and spectrum~\cite{BRODSKYPREP}. For example, this framework has been used as the conceptual basis for building a confining LF  Hamiltonian model for quarks and gluons, where the hadron eigenstates are found
 by diagonalization, resorting to the basis light-front quantization (BLFQ) method~\cite{Vary:2009gt,Vary:2016emi}. This approach  has been applied to  study  light meson  parton distribution functions~\cite{Lan:2019vui} and generalized parton distributions~\cite{Adhikari:2021jrh}.

Each hadron reflects the full complexity of QCD in Minkowski space, e.g., the LF wave function consists of an infinite sum of Fock components dynamically coupled by the interaction~\cite{bakker2014light}. However,  Haag’s theorem~\cite{Haag:1955ev,Streater:1989vi} implies that  the formulation of interacting and free quantum field theories can only be done on inequivalent Hilbert-space representations of the field algebra, which from the fundamental point of view could turn questionable the construction of the Fock-space basis from the free LF QCD Hamiltonian to describe hadron states. This issue has been recently  circumvented in~\cite{Polyzou:2021qpr} by showing the equivalence of the scattering theory formulated in instant and light-front quantum field theories.  This study was based on a two-Hilbert space representation, without demanding the existence of a free dynamics on the Hilbert space and the adopted one particle basis, which also belongs to the asymptotic states, is built from the interacting theory. On the other hand, dressed quarks and gluons are confined and rigorously can not be  asymptotic states of QCD, despite of that this work suggests  the possibility of using dressed degrees of freedom as the building
blocks of the hadron LF wave function. 

Following this reasoning, we adopt the  phenomenological perspective, where the dressed quark is considered as the effective degree of freedom, which builds, in particular, the pion state.
Such practical point of view has already been applied to study light mesons starting from Euclidean frameworks~(see e.g.~\cite{Roberts:2021nhw}). Our work add to previous studies the exploration of the Minkowski space structure of the pion, concomitantly with constraints from  DCSB. We mention that models incorporating DCSB have been indeed explored\cite{Biernat:2013aka,Biernat:2015xya}, as in~\cite{BroPLB10} where the pion Bethe-Salpeter amplitude of a chiral model was compared with that extracted from lattice QCD calculations. More recently, the pion quasi-distributions were also obtained~\cite{Broniowski:2017wbr}.
 
The pion model proposed in Ref.~\cite{Mello:2017mor}, is build from a simple and analytical quark mass function  fitted to LQCD results~\cite{ParPRD06} in the space-like momentum region (see also~\cite{Oliveira2019}). At the same time, given the Goldstone boson nature of the pion, the scalar part of the quark self-energy provides the  pseudo-scalar component of the quark-pion vertex. This is  demanded  by the axial-vector Ward-Takahashi identities in the
limit of vanishing current quark mass~(see e.g.~\cite{CloPPNP14,horn2016pion}). It is important to mention that it is not yet known how the Fock components of the pion LF wave function, beyond the valence one, are constructed in terms of dressed quarks and gluons~\cite{Arrington_2021} in a way to satisfy Haag's theorem.
 However, we note that in the case of the pion the valence accounts for 70\% of the normalization of the wave function~\cite{Paula_2021,Ydrefors:2021dwa,Paula_2022} found in a dynamical model where the quark does not have momentum dependent self-energy and it is an eigenstate of the  LF free Hamiltonian. Therefore, it could be questionable if such  a valence content is maintained once the quark is dressed. 

In the present work,  the analytical structure of the BS amplitude of the pion is further explored, using the Nakanishi integral representation (NIR)~\cite{nak63,Nakanishi:1971} considering dressed quarks. An essential aspect needed in our work is K\"all\'en-Lehmann  (KL)~\cite{itzykson2012quantum}  spectral representation of the  quark propagator. Here we derive  the Nakanishi weight functions and make an application to the pion model of Ref.~\cite{Mello:2017mor}.
Furthermore, we obtain the spin decomposition of the valence probabilities, the longitudinal and transverse momentum distributions, and  the pion decay constant.

As further observations, we mention that integral representations are useful to expand the applicability of calculations performed in Euclidean space, such as those from lattice QCD or  from continuous methods which solves the Schwinger-Dyson and Bethe-Salpeter equations, allowing to construct the Bethe-Salpeter amplitude  in Minkowski space, and with this  eventually access partonic distributions, as already done in Ref.~\cite{LeiPRL13} for the pion. The KL spectral representation has also been used to solve non-perturbatively the Schwinger-Dyson equations for the fermion self-energy in Minkowski space~\cite{Sauli:2006ba,jia2019minkowskispace,Solis_2019,Mezrag_2021,DuartePRD2022}. In addition,
the NIR has already been applied to solve the BS equation in Minkowski space,
pioneered  in Refs.~\cite{Kusaka,Kusaka:1995za,Karmanov:2005nv}. Later on, the combination with the projection onto the LF~\cite{Karmanov:2005nv} was also used to solve the BS equation for the $0^+$ bound state of two fermions~\cite{Carbonell:2010zw,dePaula:2016oct,dePaula:2017ikc} and to calculate the excited states of the Wick-Cutkosky model\cite{Pimentel:2016cpj}. Particularly, in this approach the valence LF wave function  corresponds to a Stieltjes transform~\cite{Carbonell:2017kqa,Karmanov:2021imh}, which will be extensively used in the present study  of the pion.

This work is organized as follows. In Sect.~\ref{sec:Generalframe}  the general framework is presented, with the  key ingredients, namely the KL  representation of the scalar and vector parts of the quark propagator,  the Nakanishi integral representation of the $\pi-q\bar q$ pseudoscalar vertex and the BS model amplitude. Its decomposition in an non-orthogonal and orthogonal basis in the Dirac spinor space is also done, preparing the setup for the Nakanishi integral representation of the BS amplitude, that is presented in Sect.~\ref{sec:IR}. In that section,
we introduce an auxiliary functional, which allows to
build the  weight of the Nakanishi integral representation for the scalar functions associated with the non-orthogonal and orthogonal basis  in terms of the spectral densities.
In Sect.~\ref{sec:LFWF}, in regard to the  valence LF wave function, we  present its form in terms of the weight functions,  its spin decomposition, the associated probabilities and decay constant.
In Sect.~\ref{sec:appl}, we made the quantitative application of the developed framework to  the pion Bethe-Salpeter amplitude model of Ref.~\cite{Mello:2017mor} with dressed quarks. We obtain the Nakanishi weight functions, discuss their properties and the normalization, which now has the contribution from the running quark mass. In Sect.~\ref{sec:valenceprob}, we present the results for the  valence probability and decay constant. In Sect.~\ref{Sec:LFprojampl}, we present the results for valence wave function, its spin components, obtained from the LF projected amplitudes, and the longitudinal and transverse momentum distributions. In Sect.~\ref{sec:conclusion}, the conclusions of our study are presented. One appendix is provided with details of the necessary steps for the transformation of the weight functions between  the two basis.

\section{General framework}
\label{sec:Generalframe}

In this work we calculate pion observables using a model for the Bethe-Salpeter amplitude (BSA). The dressed fermionic propagators are written in terms of K\"allen-Lehman integral representation and the BSA is described in terms of the Nakanishi integral representation~ \cite{Nakanishi:1971}.

\subsection{Key ingredients}\label{subsec:KeyIngredi}

Our goal is to describe the pion as a dressed quark-antiquark bound state. The corresponding Bethe-Salpeter amplitude is given by 
\begin{equation}
 \Psi_\pi (k;p)=S_F\left(k_q\right)\,
 \Gamma_\pi(k;p)\,
 S_F\left( k_{\overline q}\right), 
 \label{bsa}
\end{equation}
where the quark and antiquark momenta are $k_q=k+\frac{p}{2}$ and
$ k_{\overline q}=k-\frac{p}{2}$, respectively.  In this model, the dressing of the quarks is also incorporated in the quark propagators, besides the vertex function which carries all the complexity from QCD.

The pion quark-antiquark pseudo-scalar vertex, denoted here by $\Gamma_\pi(k,p)$, is 
constrained by the pseudoscalar nature of the pion:
\begin{multline}
\label{vertex}
  \Gamma_\pi (k,p) = \gamma_5 [\imath E_\pi (k,p)+ \pslash  \, F_\pi (k,p)
\\ + k^\mu p_\mu\,  \kslash\, G_\pi (k,p) + \sigma_{\mu\nu} k^\mu p^\nu H_\pi (k,p)]\, ,
\end{multline}
where $E_\pi(k,p)$, $ F_\pi (k;p) $, $ G_\pi (k;p) $ and $H_\pi (k,p)$ are four scalar amplitudes.

The dressed quark propagator is written as \cite{horn2016pion,Mello:2017mor}, 
\begin{eqnarray}
\label{sfaltern}
&& S_F(k)=\imath\,  S_v(k^2)\,\kslash \,+\,\imath\,S_s(k^2)\nonumber \\ &&
=\imath \left(A(k^ 2)\,\kslash- B(k^2)\right)^{-1}
\, .
\end{eqnarray}
The self-energies can also be written in terms of the scalar functions introduced for the propagator, namely:
\begin{eqnarray}\label{eq:ASsSv}
&&A(k^2)=\frac{S_v(k^2)}{k^2 S_v^2(k^2)-S_s^2(k^2)}\, , \\
&& B(k^2)= \frac{S_s(k^2)}{k^2 S_v^2(k^2)-S_s^2(k^2)}\, .
\label{eq:BSsSv}
\end{eqnarray}

The scalar functions $S_v(k)$, and $S_s(k)$ can be written in  terms of the
K\"allen-Lehmann spectral decomposition~\cite{itzykson2012quantum},
\begin{eqnarray}
\label{specsv}
&& S_v(k^2)=\int_0^\infty d\mu^2 \frac{\rho_v(\mu^2)}{k^2-\mu^2+\imath\varepsilon}\, ,
\\
   && S_s(k^2)=\int_0^\infty d\mu^2 \frac{\rho_s(\mu^2)}
{k^2-\mu^2+\imath\varepsilon}\, .
\label{specss}
\end{eqnarray}

Considering the chiral limit, 
$m_\pi=0$, we have for the vertex, Eq.~\eqref{vertex}, the 
expression~\cite{Mello:2017mor}
\begin{equation}
E_\pi (k;p) =\imath \,B (k^2)/f^{0}_\pi\, .
\label{vpi} 
\end{equation} 
In addition, one can also write the scalar self-energy $B(k^ 2)$ in terms of an integral representation:
\begin{equation} \label{spectralB}
B(k^2)=\int_0^\infty d\mu^2 \frac{\rho_B(\mu^2)}{k^2-\mu^2+\imath\varepsilon}
~.
\end{equation}
  We observe that, in principle by  taking the imaginary
parts of $A(k^2)$ and $B(k^2)$ in   Eqs.~\eqref{eq:ASsSv} and \eqref{eq:BSsSv} one can obtain the spectral densities for $A(k^2)$ and $B(k^2)$ in terms of $\rho_s(\mu^ 2)$ and $\rho_v(\mu^ 2)$, which for our purposes is not necessary to be detailed.

Considering only the dominant component of the vertex function ($\gamma_5$), the Bethe-Salpeter amplitude has the following
form:
\begin{multline}
	\Psi_\pi (k,p)=\gamma_5~\chi_1(k,p)+\psla{k}_q\gamma_5~\chi_2(k,p)\\ +
	\gamma_5~\psla{k}_{\overline q}~\chi_3(k,p)  +
	\psla{k}_q~\gamma_5~\psla{k}_{\overline q}~\chi_4(k,p)~\, ,
		\label{bsopenchi}
\end{multline}
where the scalar amplitudes $\chi_i(k,p)$  are defined in terms of $S_v(k^2)$, $S_s(k^2)$ and $B(k^2)$, as explicitly shown in what follows.

\subsection{ Non-orthogonal basis}
\label{subsec:nonorthbasis}

The scalar amplitudes in the pion BS amplitude in the non-orthogonal basis decomposition,
Eq.~\eqref{bsopenchi},
are written in terms of an auxiliary functional of the spectral densities as:

\begin{eqnarray}
 &&   \chi_1(k,p)=\mathcal{F}(k,p;[\rho_B,\rho_s,\rho_s])\, , \nonumber\\ 
 &&       \chi_2(k,p)=\mathcal{F}(k,p;[\rho_B,\rho_v,\rho_s])\, ,\nonumber \\ && \chi_3(k,p)=\mathcal{F}(k,p;[\rho_B,\rho_s,\rho_v])\,,   \nonumber\\ && \chi_4(k,p)=\mathcal{F}(k,p;[\rho_B,\rho_v,\rho_v])\, ,
\end{eqnarray}

where the functional is defined as
\begin{eqnarray}\label{functionalI}
    && \mathcal{F}(k,p;[o,f,h]) = {1\over f_{\pi}^{0}}
    \int_0^\infty d\mu''^2 \frac{o(\mu''^2)}{k^2-\mu''^2+\imath\varepsilon}
 \nonumber   \\
 &&\times   \int_0^\infty d\mu^{\prime 2}
		  \frac{f(\mu^{\prime 2})}{k_q^2-\mu^{\prime 2}+\imath\epsilon} 
		\int_0^\infty d{\mu}^2\frac{h({\mu}^2)}
		{k_{\overline q}^2-{\mu}^2+\imath\epsilon}\, . \nonumber\\
\end{eqnarray}
The symmetry properties of the scalar functions
\begin{eqnarray}
 &&   \chi_1(k,p)=\chi_1(-k,p)\, , \nonumber \\ 
 &&       \chi_2(k,p)=\chi_3(-k,p) \nonumber\, , \\  &&
 \chi_4(k,p)=\chi_4(-k,p)\, ,
 \label{chisym}
\end{eqnarray}
are reflected in the weight functions of the Nakanishi integral representation representing the BSA. In the following we present the decomposition of the BSA in an orthogonal basis used to study the pion with a dynamical model in Minkowski space~\cite{Paula_2021}.

\subsection{Orthogonal basis} 
\label{subsec:orthbasis}

The Bethe-Salpeter amplitude, $ \Psi_{\pi}(k,p)$, can be decomposed  as:
\begin{eqnarray}
\Psi_{\pi}(k,p) &= & \sum_{i=1}^{4}S_i(k,p)\phi_i(k,p) \, ,
\label{newbasis}
\end{eqnarray}
where the orthogonal basis for the pion $0^ -$ states is given by:
\begin{eqnarray}
S_1(k,p) & = & \gamma^5, \quad 
S_2 (k,p) =  \dfrac{\ppslash}{M_\pi}\gamma^5,
 \nonumber \\
S_3(k,p) & = &  \dfrac{k\cdot p}{M_\pi^3}~\ppslash \gamma^5~
-~\dfrac{1}{M_\pi}\kslash\gamma^5
,\nonumber \\
S_4(k,p) & = & \dfrac{\imath}{M^2_\pi} \sigma^{\mu \nu} p_{\mu} k_{\nu} \gamma^5~,
\label{elementosbase}
\end{eqnarray}
here,~$\sigma^{\mu \nu}=\frac{\imath}{2}(\gamma^{\mu}\gamma^{\nu}-
\gamma^{\nu}\gamma^{\mu})$.
 The set of operators satisfies the orthogonality condition, given by
\begin{equation}\label{ortho}
    \text{Tr}[S_i(k,p)~S_j(k,p)]~= \mathcal{N}_{i} \,  \delta_{ij}\, .
\end{equation} \, where
\begin{eqnarray}
\mathcal{N}_{1} &=& - \mathcal{N}_{2} = 4  \, , \nonumber\\
\mathcal{N}_{3} &=& \mathcal{N}_{4} = \frac{4}{ M_{\pi}^{4}} \left( (k\cdot p)^2 - M_{\pi}^2 \, k^2\right) \, .
\end{eqnarray}

The scalar functions $\phi_i(k,p)$ depends on the scalars $(k^2,p^2,k\cdot p)$, having
well defined properties under the exchange $k~\rightarrow - k$ due to its the pion fermionic nature,  namely
they have to be even for $i = 1, 2, 4$ and odd for $i = 3$ (see e.g. \cite{Carbonell:2010zw,dePaula:2016oct,dePaula:2017ikc}). 
From the orthogonality conditions, they are written as:
\begin{equation}\label{decomp}
    \phi_i(k,p)= \frac{1}{ \mathcal{N}_{i}} \,  \text{Tr}[S_i(k,p)\Psi_\pi(k,p)] \, .
\end{equation}

The scalar amplitudes $\phi_i(k,p)$  can be written in terms of $\chi_i(k,p)$ by using Eq.~\eqref{decomp} with the BSA written in Eq.~\eqref{bsopenchi}. The result is:
 \begin{eqnarray}
 \phi_1(k,p) & =& \chi_1(k,p)-\left(k^2-\frac{M_{\pi}^2}{4}\right)\chi_4(k,p)\, ,
 \label{fi1}
 \\
 \label{fi2}
 \phi_2(k,p)&= &
 \left( \frac{M_{\pi}}{2}
 +\frac{p\cdot k}{M_{\pi}}\right)\chi_2(k,p)\nonumber
 \\&+&\left(\frac{M_{\pi}}{2} -\frac{p\cdot k}{M_{\pi}}\right)\chi_3(k,p)\, ,
 \\
 \label{fi3}
 \phi_3(k,p)&=&M_{\pi}~(\chi_3(k,p)-\chi_2(k,p))\, ,
 \\
 \label{fi4}
    \phi_4(k,p)&=&M^2_\pi\,\chi_4(k,p)\, . 
 \end{eqnarray}
 Note that the symmetry properties of $\phi_i(k,p)$ under the transformation $k\to-k$ are consistent with the ones fulfilled by $\chi_i(k,p)$ expressed by \eqref{chisym}, taking into account 
the above relations.

\section{Integral Representation } \label{sec:IR}

The denominator of a generic Feynman diagram contributing to the
fermionic transition amplitudes has the same expression as
in the boson case analyzed by Nakanishi~\cite{Nakanishi:1971}. Based on this fact, we write our amplitudes following the Nakanishi integral representation:
\begin{small}
\begin{equation}
\label{eq:NIRNORTH}
\chi_i(k;p)=\int_{-1}^1\hspace{-.2cm} dz\int_{-\infty}^\infty \hspace{-.2cm} d\gamma
\frac{G_{i}(\gamma,z)}
{[k^2+z~k\cdot p-\gamma + \imath \epsilon]^3},
\end{equation}
\end{small}
One can also write the  scalar functions $\phi_i(k,p)$ with NIR as: \begin{small}
 \begin{equation}
 \label{eq:NIRORTH}
 \phi_i(k,p)=\int_{-1}^1\hspace{-.2cm} dz'\int_{-\infty}^\infty \hspace{-.2cm}  d\gamma'
 \frac{g_i(\gamma',z')}{[k^2+z'~k\cdot p-\gamma' + \imath \epsilon]^3}\, .
 \end{equation}
 \end{small}

 \subsection{Auxiliary functional}
 \label{subsec:auxfunc}
 
In order to write the scalar functions $\chi_i(k,p)$ and $\phi_i(k,p)$ in terms of NIR, we have to express the product of the three denominators in $ \mathcal{F}(k,p;[f,h,o])$ through the integral representation, which is achieved by resorting to the Feynman parametrization as derived in \cite{Mello:2017mor}, and written below:
\begin{multline}
\int_{-1}^1 dz\int_{-\infty}^\infty  d\gamma\,
\frac{\imath\,F(\gamma,z;\mu'',\mu^{\prime},\mu)}
{[k^2+z\,k\cdot p-\gamma + \imath \epsilon]^3}
=\\ 
\frac{1}{[(k+p/2)^2-\mu^{\prime 2}+\imath\epsilon][k^2-\mu''^2+\imath\epsilon]} \\
\times
\frac{1}	{[(k-p/2)^2-{\mu}^2+\imath\epsilon]}\, ,
\end{multline}
where the weight function  is given by:
	\begin{multline}
	F(\gamma,z;\,\mu'',\mu^{\prime},\mu)=\\ -
	\frac{2 \imath \,\theta(z-2~\alpha+1)~\theta(\alpha- z)~\theta(1-\alpha)~
		\theta(\alpha)}
	{|\mu^{\prime 2}+\mu^2-\frac{M_{\pi}^2}{2}-2\mu''^2|}\,,
	\label{funcaof}
	\end{multline}
with
\begin{equation}
\alpha=\frac{\gamma-z(\mu''^2-\mu^2+\frac{M_{\pi}^2}{4})-\mu''^2}
{\mu^{\prime 2}+\mu^2-\frac{M_{\pi}^2}{2}-2\mu''^2}\, .
\label{alpha}
\end{equation}

The support of the function $F(\gamma,z;\,\mu'',\mu^{\prime},\mu)$ in $\gamma$ is finite and constrained by the  theta functions for each values of the variables
$-1\leq z\leq 1$, $\mu''$, $\mu^{\prime}$ and $\mu$ for a given pion mass. One can check that $\gamma>0$ when $\mu+\mu'> M_\pi$, $\mu''+\mu'> M_\pi$ and $\mu+\mu''> M_\pi$.
In addition, the function $F$ has the symmetry property, although not immediately obvious:
\begin{equation}\label{fsym}
    F(\gamma,z;\,\mu'',\mu^{\prime},\mu)=F(\gamma,-z;\,\mu'',\mu,\mu')\, ,
\end{equation}
which corresponds to the invariance of the product of the three denominators in the right side of Eq.~\eqref{funcaof} under the transformation $k\to-k$ and  $\mu'\leftrightarrow \mu$, associated with the symmetry properties \eqref{chisym} of the $\chi_i$ amplitudes.
Another property of the weight function $F(\gamma,z;\,\mu'',\mu^{\prime},\mu)$ is that it strictly vanishes at the end-points $z=\pm 1$, which is easily checked in Eq.~\eqref{funcaof}, due to the product of the theta functions.

By using the auxiliary integral representation given in Eq.~\eqref{funcaof}, the functional $ \mathcal{F}$ of Eq.~\eqref{functionalI} is written as a NIR form:
 \begin{multline}
   \mathcal{F}(k,p;[o,f,h])=\\ =
\int_{-1}^1 dz\int_{0}^\infty  d\gamma
\frac{H(\gamma,z;[o,f,h])}
{[k^2+z'~k\cdot p-\gamma + \imath \epsilon]^3} \, ,
\end{multline}
where the weight functional
is:
\begin{multline}\label{functionH}
    H(\gamma,z;[o,f,h])=
    \int d\mu''^2 d\mu'^2 d\mu^2
   \, F(\gamma,z;\,\mu'',\mu',\mu) \\
    \times o(\mu''^2)\,f(\mu'^2)\,h(\mu^2)\, ,
\end{multline}
with the integrals having support from 0 to $\infty$. 

\subsection{Weight functions: non-orthogonal basis}\label{subsec:wfnonortho}

The weight functions of the scalar amplitudes $\chi_i$ are:

\begin{eqnarray}\label{G1H}
 &&   G_1(\gamma,z)=H(\gamma,z;[\rho_B,\rho_s,\rho_s])\, , \\ 
 \label{G2H}
 &&       G_2(\gamma,z)=H(\gamma,z;[\rho_B,\rho_v,\rho_s])\, , \\ 
 \label{G3H}
 && G_3(\gamma,z)=H(\gamma,z;[\rho_B,\rho_s,\rho_v])\,,   \\ 
 \label{G4H} && G_4(\gamma,z)=H(\gamma,z;[\rho_B,\rho_v,\rho_v])\, ,
\end{eqnarray}
and one immediately have the following symmetry properties for the weight functions taking into account Eq.~\eqref{fsym}, namely:
\begin{eqnarray}
\label{G1s}
 &&   G_1(\gamma,z)=G_1(\gamma,-z)\, , \\ 
 \label{G2G3s}
 &&       G_2(\gamma,z)=G_3(\gamma,-z)\, , \\ 
 \label{G4s} && G_4(\gamma,z)=G_4(\gamma,-z)\, ,    
\end{eqnarray}
which also implies in the symmetry properties expressed in Eq.~\eqref{chisym} 
for the scalar amplitudes.

\subsection{Weight functions: orthogonal basis}
\label{subsec:wfortho}

 Using the previous results for the functions $\phi_1(k,P)$ written in terms of the $\chi$'s with Eq.~\eqref{fi1} and the NIR from Eq.~\eqref{eq:NIRNORTH}, 
 we get:
 \begin{multline}
 \phi_1(k,p)=
 \int_{-1}^1dz \int_{0}^{\infty} d\gamma\,\Bigg\{
 \frac{G_1(\gamma,z)}
 {\left[k^2+z\,k\cdot p-\gamma' + \imath \epsilon\right]^3}
 \\
 +	\frac{\left(\frac{M_{\pi}^2}{4}-k^2\right)G_4(\gamma,z)}
 	{\left[k^2+z\,k\cdot p-\gamma + \imath \epsilon\right]^3}
 \Bigg\}\, .
 \label{confusao}
 \end{multline}

The final expression for the 
weight function is found after the $k^2$ factor is eliminated as shown in the Appendix~\ref{appendixa}, and resorting to the uniqueness of
  Nakanishi weight function, one finds that
 	\begin{multline}
	g_1(\gamma,z)=G_1(\gamma,z)+(M_{\pi}^2/4-
	\gamma)G_4(\gamma,z) \\
+\int_{0}^{\gamma} d\gamma'
\Big(G_4(\gamma',z)-
z~\partial_{z}
[G_4(\gamma',z)]\Big).
		\label{gg1}
 	\end{multline}

 The scalar amplitude $\phi_2(k,p)$  is written in terms of $\chi$'s as given by 
 Eq.~\eqref{fi2}, and their respective NIR's:
	\begin{multline}\label{g2kp} 
\phi_2(k,p)	=
\int_{-1}^1dz	\int_{0}^{\infty} d\gamma\,\Bigg\{
		\frac{\left( \frac{M_\pi}{2}+\frac{k\cdot p}{M_\pi}\right)G_2(\gamma,z)}{[k^2+z\,k\cdot p-\gamma + \imath \epsilon]^3}
		 \\
			+
		\frac{\left(\frac{M_\pi}{2}-\frac{k\cdot p}{M_\pi}\right)G_3(\gamma,z)}{[k^2+z\,k\cdot p-\gamma + \imath \epsilon]^3}\Bigg\}
\, .	\end{multline}

The final expression for the function $g_2(\gamma,z)$ is found after  the factor $k\cdot p$ in Eq.~\eqref{g2kp} is eliminated by following the same steps in deriving  Eqs.~\eqref{eq4:appa} and \eqref{eq5:appa}. The resulting weight function is, 
\begin{multline}  g_2(\gamma,z)=\frac{M_{\pi}}{2}~\left[G_2(\gamma,z)+G_3(\gamma,z)\right] \\ -\frac{2}{M_{\pi}}\int_{0}^{\gamma}d\gamma' ~ \partial_{z}\left[z\left(G_2(\gamma',z)-G_3(\gamma',z)\right)\right].	\label{gg2} 	
\end{multline}

 The third scalar amplitude $\phi_3$ is given in Eq.~\eqref{fi3}, and is written
in detail as:
 \begin{multline}
 \phi_3(k,p)
  =-\int_{-1}^1dz \int_{0}^{\infty} d\gamma\frac{M_{\pi}~G_2(\gamma,z)}
 {\left[k^2+z\,k\cdot p-\gamma + \imath \epsilon\right]^3}
\\ +
 \int_{-1}^1dz\int_{0}^{\infty} d\gamma \frac{M_{\pi}~G_3(\gamma,z)}
 {\left[k^2+z\,k\cdot p-\gamma + \imath \epsilon\right]^3}~,
 \end{multline}
 and resorting to the uniqueness of NIR, one finds that:
 \begin{equation}
 g_3(\gamma,z)=M_{\pi}~[G_3(\gamma,z)-G_2(\gamma,z)]
 .
 \label{gg3}
 \end{equation}
 
The fourth scalar amplitude $\phi_4(k,p)$ as given by Eq.~\eqref{fi4}, and NIR is written as
 \begin{multline}
 	\phi_4(k,p)=
 	\int_{-1}^1dz \int_{0}^{\infty} d\gamma~
 	\frac{ M_{\pi}^2\,G_4(\gamma,z)}{(k^2+z\,k\cdot p-\gamma + \imath \epsilon)^3} \, ,
 \end{multline}
which leads to
 \begin{equation}
 g_4(\gamma,z)~=~M_{\pi}^2~G_4(\gamma,z).
 \label{gg4}
 \end{equation}
 
 The symmetry properties of the
 $g_i$'s follows from their representation in terms of the $G_i$'s, and are given by:
 \begin{eqnarray}
\label{g124s}
 &&   g_i(\gamma,z)=g_i(\gamma,-z)\, ,\quad (i=1,2,4) \,,
 \\
 &&       \label{g3s} g_3(\gamma,z)=-g_3(\gamma,-z)\, .   
\end{eqnarray}
To obtain the above relations,  the symmetry properties of the $G_i$'s
from Eqs.~\eqref{G1s}-\eqref{G4s} were used in  Eqs.~\eqref{gg1}, \eqref{gg2}, 
\eqref{gg3} and \eqref{gg4}, which give the explicit expressions of the $g_i$'s in terms of 
linear combinations of $G_i$'s.

\subsection{Normalization}\label{sec:norm}

In order to calculate the observables of the pion, which in our case are the
valence probability, moment distributions and the decay constant, the BS amplitude must be properly normalized. We will assume that the BS amplitude of the QCD-inspired pion model is hypothetically a solution of a BS equation whose interaction kernel depends only on the transferred moment, and the
quark propagators are dressed, so the normalization expression reduces to (see~\cite{Lurie}):
\begin{small}
\begin{multline}
Tr \Bigg[\int \frac{d^4k}{(2 \pi)^4}
\frac{\partial}{\partial p^{\prime\mu}}
 \{ S^{-1} (k-p'/2)
 \bar\Psi_\pi(k,p)
 \\
 \times
 S^{-1} (k+p'/2) \Psi_\pi(k,p) \}
 \Bigg]_{p=p'} = -\imath ~ 2 p_{\mu} \, .
\label{nap1b}
\end{multline}
\end{small}
We observe that within the ladder approximation of the BS equation, Eq.~(\ref{nap1b})  also provides the pion charge normalization \cite{Ydrefors:2021dwa}.

The most general expressions for the BSA of a $0^{-}$ system is given in Eq.~\eqref{newbasis} for the ortoghonal basis, and the associated  conjugate is,
\begin{eqnarray}
&&\bar\Psi_\pi(k,p)=-S_1(k,p) \, \phi_1(k,p) + S_2 (k,p)\,  \phi_2(k,p) \nonumber \\
&+& S_3(k,p) \, \phi_3(k,p) + S_4(k,p) \, \phi_4(k,p) \, , 
\label{conjugate amplitude}
\end{eqnarray}
where $S_i(k,p)$ are the Dirac operators from Eq.~\eqref{elementosbase}.  We have already defined these scalar functions, $\phi_i$(k,p), which contain the analytic behavior imposed by the Feynman prescription, namely $+\imath \epsilon$. These have to obey well-defined properties under the $k \to -k$ exchange, according to the anticommutation rule for the fermionic fields.

A comment is appropriate here. The normalization formula in
Eq.~\eqref{nend1}, contains  contributions from the LF Fock-space decomposition of the pion wave function beyond the valence one, that is, it takes into account the infinite sum of states with the dressed quark-antiquark pair and any number of dressed gluons, which have the pion quantum numbers. All this complexity appears when the Bethe-Salpeter equation is projected onto the light-front, and Fock components beyond the valence one  contribute to the kernel of the effective squared mass operator that acts on the valence component~\cite{Sales:1999ec}. Also the BS amplitude as defined in Minkowski space, has  the quark and antiquark propagating in different LF times, and involves from the point of view of the LF Hamiltonian evolution a propagation in the relative time, and therefore during the virtual propagation between these two times the quantum system fluctuates on all Fock components
 allowed by the pion quantum numbers. Thus, although the  $\pi-q\bar q$ BS amplitude   is defined in terms of the matrix element of only two quark operators, evaluated between the vacuum and the pion states, the quantization of the quark and gluon fields and the non-diagonal form in the  evolution operator in the LF Fock-space end-up allowing the virtual propagation of the system in all Fock components allowed by the pion quantum numbers. There remains the difficult question on how the LF evolution operator can be defined for the dressed degrees of freedom.

\section{LF valence wave function} \label{sec:LFWF}

The valence wave function is the Fock space component of the hadron state with the smallest number of constituents that carries its quantum numbers. The hadron state on the null-plane is decomposed in an infinite sum of Fock-components with an arbitrary number of dressed constituents, namely quarks and gluons, and each of these many-body state carrying the hadron quantum number. All components contribute to the normalization and, in particular, to the valence probability. In the pion it  is associated to the probability to find a dressed pair  of quark and an antiquark. In the most simplest case of constituent quarks with no structure, the basis in Fock space can be formulated using the standard methods of quantum field theory on the Light-Front, where one
defines the creation and annihilation operators for particles and antiparticles in the null plane with arbitrary spin, and the hadron state is decomposed in Fock-components with a generic number of constituents~\cite{BRODSKYPREP}. 

On the other hand, it is recognized that the valence wave function comes from the  elimination of the relative LF time between the minimum number of quark operators that enter the matrix element between the vacuum and the hadron state, and that defines the BS amplitude ~\cite{Sales:1999ec,Frederico:2010zh}. Alternatively, the valence wave function can be obtained using the quasi-potential expansion method adapted to perform the projection on the light-front
of the BS equation and associated amplitude~\cite{Sales:1999ec,Frederico:2010zh}. It is worthwhile to mention that even the BS amplitude
with the minimal number of legs contains the information on the full
LF Fock-space composition of the hadron. It should be noted that gauge links are always required between the
quark field operators to define an observable in order to keep color gauge invariance~\cite{Collins:2011zzd}. Recently, it was suggested~\cite{XJIRMP2021} to obtain the LF probability amplitude of the hadron many-body component from the matrix elements between the vacuum and the hadron state of quark and gluon operators connected by  gauge links, and then eliminate the relative LF time by projecting these amplitudes onto the null-plane.

\subsection{Spin components}\label{subsec:spin}

 In our phenomenological treatment of the pion state, the formulation of the valence wave function with dressed constituents will follow Ref.~\cite{Paula_2021}, which derived the relation between the valence components with antialigned and aligned quark spins in terms of the scalar functions associated with the decomposition of the BSA in the orthogonal Dirac basis:
\begin{small}
\begin{multline}
\varphi_2(\xi,\vec{k_{\perp}},\sigma_i,J^{\pi},J_z)=
-\frac{\sigma_2}{2}\sqrt{N_c}\sqrt{1-z^2}
\\
\times \int\frac{dk^-}{2\pi}
\Bigg\{\delta_{\sigma_2,-\sigma_1} \left(\phi_2(k,p)+\left( \frac{k^-}{2M_{\pi}}+\frac{z}{4 }\right)\phi_3(k,p)\right)
\\
-\frac{k_{L(R)}\sqrt{2}}{M_{\pi}}~\delta_{\sigma_2,\sigma_1}~\phi_4(k,p) \Bigg\} \, ,
\label{varphik-}
\end{multline}
\end{small}
for $\sigma_1=\pm 1$, $\sigma_2=\pm 1$ and $ N_{c}=3$ is the number of colors. The momentum fraction is $0<\xi<1$ and the associated relation $z=2\xi-1$. Although, strictly speaking the derivation  in~\cite{Paula_2021} was done with a structureless constituent quark degrees of freedom, we will assume the validity of this formula for dressed quarks, noting that Eq.~\eqref{varphik-} has no trace of the particular constituent quark mass, which would forbid its applicability to dressed quarks. Another indication of the utility of Eq.~\eqref{varphik-} in a general case comes from the decay constant formula, which can be written in terms of the antialigned spin component or solely from $\phi_2$~ \cite{Paula_2021}.

From the above reasoning, we  write the valence probability as given by:
\begin{multline}\label{eq:pval}
    P_{val}=\frac{1}{(2\pi)^3}\sum_{\sigma_1,\sigma_2}\int^1_{-1}\frac{dz}{1-z^2}\int d^2k_\perp \\ \times|\varphi_2(\xi,\vec{k_{\perp}},\sigma_i,J^{\pi},J_z)|^2\, ,
\end{multline} 
where we identify the two-spin components of the pion valence wave function following Ref.~\cite{Paula_2021} and written as:
\begin{small}
\begin{multline}
\varphi_2(\xi,\vec{k_{\perp}},\sigma_i;M_\pi,J^{\pi},J_z)=
-\dfrac{\sigma_2}{2}\sqrt{N_c}\sqrt{1-z^2}
\\
\times
\left\{\delta_{\sigma_2,-\sigma_1}~\Psi_{\uparrow\downarrow}(\gamma,z)~\mp~\delta_{\sigma_2,\sigma_1}~e^{\mp \imath \theta}~\Psi_{\uparrow\uparrow}(\gamma,z)
\right\}\, ,
\label{amplvalfinal}
\end{multline}
\end{small}
where $k_{x}\mp \imath \, k_y = \sqrt{\gamma} \, e^{\mp \imath \, \theta}$. In terms of the LF projected amplitudes, which were shown for the pion QCD inspired model in Sect.~\ref{Sec:LFprojampl},  the two spin components of the pion valence wave function are given  by~\cite{Paula_2021}:
\begin{small}
\begin{multline}
\Psi_{\uparrow\downarrow}(\gamma,z)=\psi_{2}(\gamma,z)+\frac{z}{2}\psi_{3}(\gamma,z)
\\
+
\frac{\imath}{M_{\pi}^3}\int_{0}^{\infty}d\gamma'
\frac{\frac{\partial}{\partial z}[g_3(\gamma',z)]}
{\left(\gamma+\gamma'+\frac{z^2M_{\pi}^2}{4} - \imath \epsilon\right)}\, ,
\label{eq:psiantipar}
\end{multline}
\end{small}
for the anti-aligned quark and antiquark spins, and
\begin{equation}\label{eq:psipar}
\Psi_{\uparrow\uparrow}(\gamma,z)=\frac{\sqrt{\gamma}}{M_{\pi}}~ \psi_4(\gamma,z)
\end{equation}
for  the aligned quark spins.

In the above expressions for the two spin components of the pion wave function, the LF time projected amplitudes are~\cite{dePaula:2017ikc,Paula_2021},
 \begin{equation}
 \psi_i(\gamma,z)=
 \int\frac{dk^-}{2\pi}~\phi_i(k,p),
 \label{wfdef}
 \end{equation}
with the notation~${\vec k}_{\perp }^2=\gamma\geq 0$ and $z=-2k^+/p^+$, where the scalar functions are represented by the NIR
written in Eq.~\eqref{eq:NIRORTH}.
The momentum are transformed to the light-front ones:
$k^-=k^0-k^3$, $k^+=k^0+k^3$ and $\vec k_\perp$. The final expression for  $\psi_i(\gamma,z)$~\cite{karmanov2006bethe},
 \begin{equation}
 \psi_i(\gamma,z)=-\frac{\imath}{M_{\pi}}
 \int_{0}^{\infty}\hspace{-.2cm} d\gamma' ~
 \frac{g_i(\gamma',z)}{\left[\frac{z^2M_{\pi}^2}{4}+\gamma
 	+\gamma'\right]^2}\,,
 \label{wfi}
 \end{equation}
 where we have used the property of the support of the Nakanishi weight functions. Therefore, the denominator never vanishes for the pion bound state, and the $ \imath \epsilon$ is irrelevant in this case.

 \subsection{Valence probability decomposition}\label{subsec:valpro}
 
 The components of the valence wave function with spins of antialigned  \eqref{eq:psiantipar} and
aligned quarks \eqref{eq:psipar} are the probability amplitudes, as long as the full wave function in Fock space is normalized. The valence probability, defined in Eq.~\eqref{eq:pval}, can be written as:
\begin{equation}
P_{\text{val}} =
 \int_{-1}^{1} dz \int_{0}^{\infty} d\gamma~
 \rho_{val}(\gamma,z)\, ,
\label{eq:pval0}
\end{equation}
where the probability densities associated with the two spin components take place:
\begin{equation}\label{eq:rhoval}
\rho_{val}(\gamma,z)=\rho_{\uparrow\downarrow}(\gamma,z)
+\rho_{\uparrow\uparrow}(\gamma,z)
\end{equation}
and are defined as
\begin{eqnarray}\label{eq:pvalud}
&&\rho_{\uparrow\downarrow}(\gamma,z)=\frac{N_c}{16\pi^2}
   |{\psi}_{\uparrow\downarrow}(\gamma,z)|^2\, , \\
&&   \rho_{\uparrow\uparrow}(\gamma,z)=\frac{N_c}{16\pi^2}\frac{\gamma}{M_{\pi}^2}|\psi_4(\gamma,z )|^2 \, ,
 \label{eq:pvaluu}
\end{eqnarray}
being the moment probability densities for the anti-aligned and aligned spin components, respectively. The probability for each valence spin component is therefore given by:
\begin{eqnarray}\label{eq:Pupdown}
    P_{\uparrow\downarrow}&=&\int_{-1}^{1} dz \int_{0}^{\infty} d\gamma~
 \rho_{\uparrow\downarrow}(\gamma,z)\, ,
 \\ P_{\uparrow\uparrow}&=&\int_{-1}^{1} dz \int_{0}^{\infty} d\gamma~
 \rho_{\uparrow\uparrow}(\gamma,z) \, ,
 \label{eq:Pupup}
\end{eqnarray}
 where we have explicitly written the two contributions to the valence state.

It is important to note that, according to~\cite{Paula_2021}, the configuration of anti-aligned quark and antiquark spins have probabilities around $P_{\uparrow\downarrow}\sim 0.55$  and $P_{\uparrow\uparrow}\sim 0.15$, respectively. Reminding that for the anti-aligned spin configuration  the quark and antiquark state is an eigenstate of the operator $L_z$ with eigenvalue $\ell_z = 0$. In contrast, the aligned configuration has the quark spins  necessarily coupled in an eigenstate of $L_z$ with $\ell_z = \pm 1$. We cannot rule out the aligned spin component, even with its low influence on the relativistic dynamical regime of the quarks within the pion.
A quantitative study of this component has a
key role in understanding the characteristics of light mesons (and possible relativistic corrections
for heavier ones).

\subsection{Decay constant}\label{sec:fpi}
 
The decay constant
is defined in terms of the BSA, as follows
\begin{equation}
\imath~p^\mu f_{\pi} = N_{c} \int \frac{d^4 k}{ (2\pi)^4}~\mbox{Tr} [\, \gamma^\mu \, \gamma^5 \, \Psi_\pi(k ,p) ] \, .
\label{fpi0}
\end{equation}
Performing the contraction with $p_\mu$, from the right in each member and using the BS amplitude decomposition, given in Eq.~\eqref{newbasis}, we will have
\begin{equation}\label{fpitrace}
\imath \, M^2_{\pi} f_{\pi} = - 4~M_\pi~N_{c}\int \frac{d^4 k}{ (2\pi)^4} ~\phi_{2} (k,p).
\end{equation}
Here we use the light front variables, remembering that when integrating $\phi_2$ into the variable that corresponds to energy, $k^-$, we obtain the component $\psi_2$, i.e.
\begin{small}
\begin{eqnarray}
\imath~M_\pi~f_{\pi} =-4~N_c~\frac{1}{2}
                  \int\frac{d^2 {k_\perp}}{(2\pi)^2}
                  \int\frac{dk^+}{(2\pi)}~\psi_2(\gamma,z) \, .
\end{eqnarray}
\end{small}
We must analyze the integrals in the longitudinal momentum, $k^+$ and $\vec k_{\perp}$ by writing $dk^+ = - \frac{M _ {\pi}}{2}dz$ and  $d^2 {k_{\perp}} =d\gamma\frac{d\theta}{2}$, where we use that $ \gamma= |\vec{k_{\perp}}|^2$, with $\theta$ in the range $[0,2\pi]$, and using the definition of $\psi_2(\gamma,z ) $, which was presented in Eq.~\eqref{wfi}:
\begin{small}
\begin{equation}
f_{\pi} =\frac{N_c}{8\pi^2M_{\pi}}
\int_{0}^{\infty} \hspace{-.2cm}d\gamma \int_{-1}^1 \hspace{-.2cm} dz
\int_{0}^\infty\hspace{-.2cm}d\gamma'
\frac{g_2(\gamma',z)}
{\left[\gamma+\gamma'+ z^2  \dfrac{M^2_\pi}{4}\right]^2},
\end{equation}
\end{small}
and finally, we just integrate in $\gamma$ and apply the limits, obtaining:
\begin{small}
\begin{equation}
f_{\pi} =\frac{N_c}{8\pi^2~M_{\pi}}
\int_{-1}^1 dz
\int_{0}^{\infty}d\gamma'
\frac{g_2(\gamma',z)}
{\gamma' + z^{2} \, \dfrac{M^2_\pi}{4}} .
\label{finalfpi}
\end{equation}
\end{small}

We stress that all observables are calculated after obtaining the normalization constant $\mathcal{N}$, present in the Nakanishi weight functions, which is obtained from the normalization condition given in Eq.~\eqref{nend1}.

We should also point out that it is possible to express ~$f_\pi$ in terms of the anti-aligned spin component of the valence wave function, given by Eq.~\eqref{eq:psiantipar}, as discussed in Ref.~\cite{Paula_2021}.
The demonstration of this equivalence was done using the ``+'' component of the axial current, and we get the expression below,
\begin{eqnarray}
f_{\pi} = \imath \frac{N_c}{8\pi^2}
\int_0^{\infty}d\gamma\int_{-1}^1 dz~\Psi_{\uparrow\downarrow}(\gamma,z)\, ,
\label{decayFL}
\end{eqnarray}
which is equivalent to the final form found in Eq.~\eqref{finalfpi}.

\section{Application}\label{sec:appl}

The present example extends the investigation of the pion  by exploring the integral representation of the BS amplitude, as developed in Sect.~\ref{sec:IR}, applying the formalism to the Minkowski space QCD-inspired model  proposed in Ref.~\cite{Mello:2017mor}.
In this model the pion Bethe-Salpeter amplitude incorporates the dressed quark propagator. Even though  only the quark momentum-dependent running mass is considered,  the decay constant and the electromagnetic form factor experimental  are reproduced~\cite{Mello:2017mor}.

The pseudo-scalar pion-quark vertex in the QCD-inspired model is proportional to the running mass of the dressed quark propagator (see Eqs.~\eqref{vpi}  and \eqref{bsopenchi}). The analytical form of the running mass contains a single pole in the time-like  region, which is fitted to the  Landau gauge  lattice QCD calculations in the Euclidean space~\cite{ParPRD06}. Such parametrization  allows to analytically extend  the BS amplitude to the Minkowski space, and therefore a more detailed study of this model exploring the  Nakanishi integral representation~\cite{Nakanishi:1971}.

In this section,  we derive the explicit form of the NIR  weight functions of the four scalar amplitudes associated with the decomposition of the pion BS amplitude using two choices of operator basis in the Dirac spinor space: (i) a  non-orthogonal basis (see Sect.~\ref{subsec:wfnonortho}), and (ii) an orthogonal basis (see Sect.~\ref{subsec:wfortho}) , where the latter one has been used to solve the BS equation in Minkowski space for the pion~\cite{Paula_2021,Ydrefors:2021dwa}. In  addition, we present the numerical results for the weight functions and for the projected scalar amplitudes on the light-front in the orthogonal basis, which  are necessary to build the valence wave function, as detailed in Sec.~\ref{Sec:LFprojampl}. The numerical results will be discussed, emphasizing their symmetry properties, as well as the characteristics brought by the running dressed quark mass.

\subsection{Model parametrization}\label{subsec:model}

The general form of the dressed quark propagator is written in the form:
\begin{equation}
\label{sfO}
S_F(k)=\imath\,Z(k^2)
\left[\kslash-M(k^2)+\imath\epsilon\right]^{-1} \, .
\end{equation}
In the adopted model from Ref.~\cite{Mello:2017mor}, 
it is used the simplification that $Z(k^2)=1/A(k^2)=1$. Such simplification 
makes the BS amplitude, in the infrared region, about twice larger than it
should be, considering that $Z(0)\sim 0.7$~\cite{ParPRD06}. On the other 
hand, such enhancement is compensated by the computed values of $f_\pi$, 
the pion charge radius and the electromagnetic form factor in close agreement with the experimental data, as shown 
in~ \cite{Mello:2017mor}.

\begin{figure}[thb]
\begin{center}

\vspace{0.5cm}

\epsfig{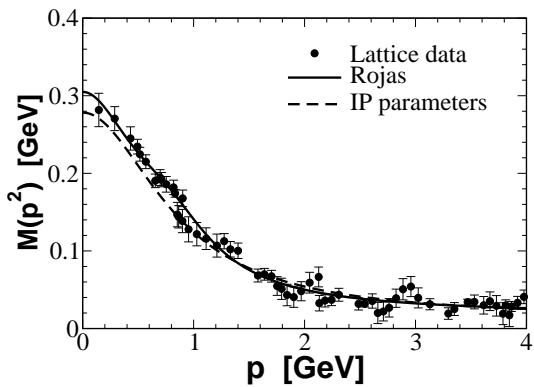}
\caption{Quark running mass model (dashed line),
 with the present parameters, $m_0=0.014$~GeV,  $m=0.574$~GeV and
$\lambda=0.846$~GeV~\cite{ Mello:2017mor};
compared  with lattice QCD calculations~\cite{ParPRD06} (full circles) and the parametrization (solid line) given in \cite{Rojas2013}. 
Results are presented for space-like momentum. } 
\label{massrunning}
\end{center}
\end{figure}

Here, in order to illustrate the general framework we have developed in the previous sections, we make use of such quark propagator model, which has the simplified form:
\begin{equation}\label{sf}
S_F(k)=\imath \frac{\kslash + M(k^2)} { k^2-M^2(k^2)+\imath\epsilon }.
\end{equation}
Following the previous work~\cite{Mello:2017mor},
we adopt a quark mass function with a monopole  shape having a single time-like pole added to the current quark  mass (see also~\cite{DUDAL2016}), which is fitted to lattice QCD calculations~\cite{ParPRD06}, and given by:
\begin{equation}
M(k^2)=m_0-m^3\left[k^2- \lambda^2 + \imath \epsilon \right]^{-1},
\label{runningmass}
\end{equation}
where
$m_0=0.014$~GeV,  $m=0.574$~GeV and
$\lambda=0.846$~GeV. 
The result of the dressed quark mass employed here is shown in Fig.~\eqref{massrunning}. It is compared to the 
lattice QCD results in the Landau gauge from Ref.~\cite{ParPRD06}, and also with the parametrization given in~\cite{Rojas2013}.

The introduction of Eq.~\eqref{runningmass} in the quark propagator presents poles in time-like region, which are the zeros of:
\begin{equation}
k^2\left(k^2-\lambda^2\right)^2-\left[m_0\left(k^2-\lambda^2\right)-m^3\right]^2~=0 \, ,
 \label{funcaomassa}
\end{equation}
and, with the given parameters, the solutions are such that only $k^2=m_i^ 2>0$ are found, with the following values:
$m_1=323$~MeV, $m_2=645$~MeV, and $m_3= 954$~MeV.

The time-like poles of the quark propagator can be factorized in the form:
\begin{equation}\label{sf1}
 S_F(k)=\imath\,\,\frac{\left(k^2- \lambda^2\right)^2\, (\kslash +m_0)- \left(k^2- \lambda^2\right)\,
m^3}
{{\prod_{i=1,3}}(k^2-m^2_i+\imath \epsilon)} \, ,
\end{equation}
and comparing with Eq.~\eqref{sfaltern}, we single out its vector and scalar components, respectively as:
\begin{equation}\label{spoles}
\begin{aligned}
 S_v(k^2) &
 =\sum_{i=1}^3
\frac{D_i}{\left(k^2-m_i^2+\imath \epsilon\right)} \, ,
\\  
S_s(k^2) 
&=\sum_{i=1}^3 \frac{E_i}{\left(k^2-m_i^2+\imath \epsilon\right)}+\,S_v(k^2)m_0
\, , 
\end{aligned}
\end{equation}
where,
\begin{equation}\label{Dv}
D_i=\dfrac{(\lambda^2-m_i^2)^2}{(m_i^2-m_j^2)(m_i^2-m_k^2)}\, ,
\end{equation}
and
\begin{equation}\label{Es}
    E_i=  - \frac{(m_i^2-\lambda^ 2)\,m^3}{(m_i^2-m_j^2)(m_i^2-m_k^2)}\, .
\end{equation}
with cyclic permutations of $\{i,j,k\}\equiv\{1,2,3\}$.

\begin{table}[]

    \centering
    
        \caption{Residue of the vector and scalar  parts given in columns three and four, respectively, of the dressed quark propagator at each pole of mass $m_i$ shown in column two.}
        
    \vspace{0.5cm}

{    
    \begin{tabular}{|c|c|c|c|}
     \hline
    $i$ &  $m_i$ [MeV] & $D_i$  & $C_i$ [GeV] \\
         \hline
   1  & 323   & 1.487 & 0.481\\ 
          \hline
   2  & 645    & -0.583  & -0.376 \\
          \hline
   3  &  954  & 0.096 & -0.091\\
          \hline
    \end{tabular}
    \label{tableresidue}
}
\end{table}

The spectral density of the vector component of the quark propagator $S_v(k^2)$ in the form of Eq.~\eqref{specsv} is
\begin{equation}
    \label{rhov}
 \rho_v(\mu^2)~=\sum_{i=1}^3D_i\,\delta(\mu^2-m_i^2)\, ,
\end{equation}
and the spectral density of the scalar part of the quark propagator in the form of Eq.~\eqref{specss} is written as: 
\begin{equation}\label{rhos}
  \rho_s(\mu^2)~=~\sum_{i=1}^ 3 C_i\,\delta(\mu^2-m_i^2)\, ,
\end{equation}
where
$C_i=E_i + m_0 \, D_i$.

In Table~\ref{tableresidue}, we present
the residue
of the poles in the scalar and vector components of the dressed quark propagator, which are weights of the Dirac-deltas in the spectral densities.  It was shown in~\cite{Mello:2017mor} that this model violates the spectral density positivity relations~\cite{itzykson2012quantum}, which are not in principle required to be satisfied, as the quark is not a colorless asymptotic state of QCD.

\subsection{Bethe-Salpeter amplitude model}\label{subsec:BSmodel}

The pion Bethe-Salpeter amplitude  is built in terms of the  quark propagator, Eq.~\eqref{sf1}, which has the scalar and vector components  given by Eq.~\eqref{spoles} with the corresponding spectral densities written in Eqs.~\eqref{rhov} and \eqref{rhos}, respectively. It reads
\begin{multline}
	\label{BSA1}
	\Psi_\pi (k;p)=- \imath \left[S_v(k_q^2)\,\psla{k}_q+S_s(k_q^2)\right]
	\, \\ \times\frac{\mathcal{N}\,\gamma_5~m^3}{k^2-\lambda^2+\imath\epsilon}
	\left[S_v(k_{\overline q}^2)\,\psla{k}_{\overline q}+S_s(k_{\overline q}^2)\right]~, 
\end{multline}
where $k_q=(k+p/2)$, and $k_{\overline q}=(k-p/2)$. The above BS amplitude fits into the form 
adopted in Eq.~\eqref{bsopenchi}, albeit with  spectral densities associated with a finite number of time-like poles in the self-energy and associated quark propagator. We should clarify that in our example,   as suggested by the model of Ref.~\cite{Mello:2017mor}, the pion vertex is obtained from the running mass term in the chiral limit when $m_0\to  0$,  while the quark propagators in the pion BS amplitude kept the finite value of $m_0$ in $M(k^2)$ as written in Eq.~\eqref{runningmass}.

The spectral density of the
vertex function is  
\begin{equation}\label{rhob}
\rho_B(\mu^2)=\mathcal{N}m^3\delta(\mu^2-\lambda^2)\, ,
\end{equation}
and with this formula we have all the ingredients to derive the Nakanishi integral representation of the four scalar amplitudes appearing in the decomposition of the BS amplitude in the Dirac spinor space. 

\begin{figure}[thb]

\begin{center}
	
\vspace{0.5cm}

\epsfig{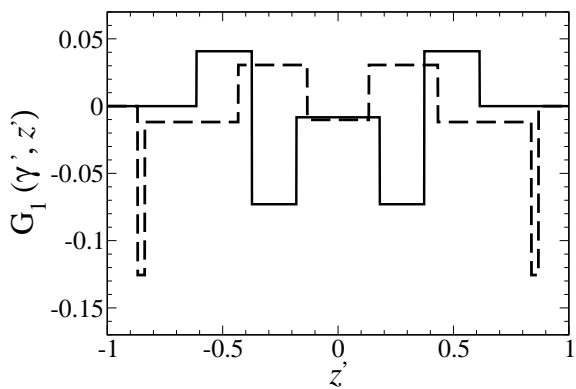} 
	\caption{$G_1(\gamma,z)$ as a function of $z$ with 
	~$\gamma=$0.75~GeV$^2$ (solid line) and 0.45~GeV$^2$ (dashed line). For  the plot we use in Eq.~\eqref{G1} the arbitrary value of $\mathcal{N}=100$, drop out the factor of $i$ and all parameters are in units of GeV.}
		\label{Fig:G1}
	\end{center} 
\end{figure}

In what follows, we will obtain the  weight functions $G_i(\gamma,z)$, associated with the  NIR of the scalar amplitudes, Eq.~\eqref{eq:NIRNORTH}, appearing in the decomposition of the BS amplitude in the non-orthogonal basis. We also compute  the weights, $g_i(\gamma,z)$, associated with the NIR of  the scalar amplitudes, Eq.~\eqref{eq:NIRORTH}, corresponding to  the orthogonal basis decomposition of the BS amplitude.

\subsection{ NIR: non-orthogonal basis}\label{subsect:NIRNORTHO}

 Having at hand the spectral densities of the vector and scalar parts of the dressed quark propagator and the one for the vertex function, Eqs.~\eqref{rhov},
 \eqref{rhos} and
\eqref{rhob}, respectively,  the NIR weight functions can be derived. The $G_i(\gamma,z)$ $(i=1-4)$ are obtained  using the general form given in Sect.~\ref{subsec:wfnonortho} and written in Eqs.~\eqref{G1H}-\eqref{G4H}. 

We remind that the auxiliary function $F(\gamma,z;\lambda,\mu^{\prime},\mu)$, defined in Eq.~\eqref{funcaof}, is essential to build the  weight functions for the scalar amplitudes $\chi_i(k,p)$ in Eq.~\eqref{eq:NIRNORTH}. The support of  the function $F(\gamma,z;\lambda,\mu^{\prime},\mu)$ in $\gamma$ with $-1\leq z\leq 1$, for
$\lambda=$~0.846 GeV, $\mu'=m_i$ and $\mu=m_j$ corresponding to the time-like poles of the dressed quark propagator, are limited by the theta's 
 in Eq.~\eqref{funcaof},
 which  strictly imposes $\gamma>0$ for each  pair $(m_i,m_j)$. Interesting to observe that   $\gamma$  has an upper bound of 1.394~GeV$^2$, meaning that the weight functions incorporate  a scale of about 1~GeV brought by the the running quark mass function parameter, $\lambda$, which was obtained by a fitting to LQCD results.

The weight function $G_1(\gamma,z)$ is calculated with Eq.~\eqref{G1H} by taking into account the spectral densities of the dressed quark propagator vector and scalar parts, and the spectral density of the pion vertex function. Therefore, by substituting Eqs.~\eqref{rhov},
 \eqref{rhos} and
\eqref{rhob}  in Eq.~\eqref{G1H},  it is found that:
\begin{equation}
G_1(\gamma,z)=w
	\hspace{-.2cm}\sum\limits_{1\leq i,j\leq 3}\hspace{-.2cm}
	C_iC_jF(\gamma,z;\lambda,m_i,m_j),
\label{G1}
\end{equation}
where $w=-\mathcal{N}m^3$ gives the proper normalization to the residue of the scalar part of the quark propagator. 
In Fig.~\eqref{Fig:G1}, we present the  weight 
function $G_1(\gamma,z)$ for two values of $\gamma$ equal to 0.45 
and 0.75~GeV$^2$, we confirm  the even property under the transformation $z\to-z$ corresponding to Eq.~\eqref{G1s}. Furthermore, one observes 
the characteristic plateaus in the weight function from the theta functions  in $F(\gamma,z;\lambda,m_i,m_j)$.

\begin{figure}[thb]
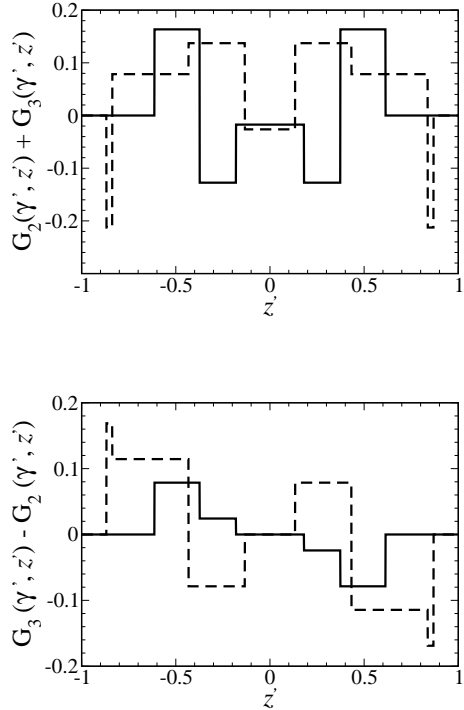

	\begin{center}
	
\vspace{0.5cm}

		\epsfig{figure=somaG2+G3.eps,width=6.cm ,angle=0} 
		
		\vspace{1cm}
		
		\epsfig{figure=diferenG2-G3.eps,width=6.cm,angle=0} 
\caption{Top panel: $G_2(\gamma,z)+G_3(\gamma,z)$ as a function of $z$
with ~$\gamma=$ 0.75~GeV$^2$ (dashed line) and 0.45~GeV$^2$ (solid line).
Bottom panel:
		~$G_3(\gamma,z)-G_2(\gamma,z)$  as a function of $z$
		with ~$\gamma=$ 0.75~GeV$^2$ (solid line) and 0.45~GeV$^2$ (dashed line). 
		For  the plot we use in Eq.~\eqref{G2} the arbitrary value of $\mathcal{N}=100$, drop out the factor of $i$ and all parameters are in units of GeV.	}
	\label{Fig:G2G3}
	\end{center} 
\end{figure}

The weight function $G_2(\gamma,z)$ is derived following the same steps that leads to $G_1(\gamma,z)$, namely we substitute the spectral densities from Eqs.~\eqref{rhov},
 \eqref{rhos} and
\eqref{rhob}  in Eq.~\eqref{G2H}, resulting in:
\begin{multline}
G_2(\gamma,z)=w
\hspace{-.2cm}\sum\limits_{1\leq i,j\leq 3}\hspace{-.2cm}
D_iC_j F(\gamma,z;\lambda,m_i,m_j).
\label{G2}
\end{multline}
The transformation property of the original form  of $G_2$ and $G_3$ from \eqref{G2H} and \eqref{G3H}  turns into the symmetry property in Eq.~\eqref{G2G3s}  gives that   $G_3(\gamma,z) = G_2(\gamma,-z)$.
Motivated by this property, we present in Fig.~\eqref{Fig:G2G3}, the even and odd combinations  $G_3(\gamma,z)+G_2(\gamma,z)$ 
and $G_3(\gamma,z)-G_2(\gamma,z)$ 
for two values of $\gamma$ equal to 0.45 
and 0.75~GeV$^2$. The plateaus from the theta functions  in $F(\gamma,z;\lambda,m_i,m_j)$  are clearly seen in the figure. Comparing to  $G_1$  the two  combinations have a factor of about $\sim 3$ smaller, which comes from the residue $D_i$ compared to $E_i+m_0D_i$ as shown in Table~\eqref{tableresidue}.
The model also presents $G_3(\gamma,z)+G_2(\gamma,z)\sim 3 G_1(\gamma,z)$. The $z$-odd function $G_3(\gamma,z)-G_2(\gamma,z)$ is also shown in the lower panel of the figure. One observe that the signs of the plateaus alternate by changing $\gamma$, as a consequence of the alternate signs of the residue (see Table~\ref{tableresidue}). 

\begin{figure}[thb]
	\begin{center}
\vspace{0.5cm}
\epsfig{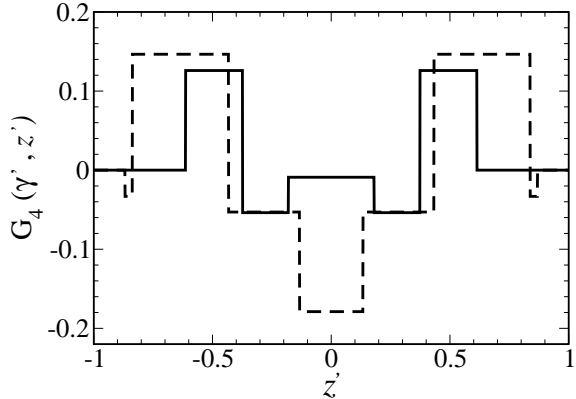} 
		\caption{ 
$G_4(\gamma,z)$ as a function of $z$	with ~$\gamma=$ 0.75~GeV$^2$ (solid line) and 0.45~GeV$^2$ (dashed line). 	For  the plot we use in Eq.~\eqref{G4} the arbitrary value of $\mathcal{N}=100$, drop out the factor of $i$ and all parameters are in units of GeV. }
		\label{Fig:G4}
	\end{center}
\end{figure}

The  weight function $G_4(\gamma,z)$ is derived by introducing the spectral densities~\eqref{rhov},
 \eqref{rhos} and
\eqref{rhob}  in Eq.~\eqref{G4H}, which results in:
\begin{equation}
G_4(\gamma,z)=w\hspace{-.2cm}
	\sum\limits_{1\leq i,j\leq 3}
\hspace{-.2cm}	D_iD_jF(\gamma,z;\lambda, m_i,m_j )\,.
\label{G4}
\end{equation}
In  
Fig.~(\ref{Fig:G4}) the dependence of $G_4(\gamma,z)$ with $z$ is shown for  $\gamma$ with values of 0.45 and 0.75~GeV$^2$. Its even property with the transformation $z\to-z$ is clearly seen, as given by the general form of the model, Eq.~\eqref{G4s}, and discussed in Sect.~\ref{subsec:wfnonortho}. Such symmetry property, although not trivially apparent when one inspects the weight function $F(\gamma,z;\lambda, m_i,m_j )$ written in Eq.~\eqref{funcaof}, leads to the symmetry properties we have seen so far in this example, is traced back to Eqs.~\eqref{chisym}, and  followed by our construction of the Nakanishi weight functions.

As an observation, the dependence of the weight functions $G_i(\gamma,z)$ in $\gamma$ for a given $z$ also exhibit the characteristic plateaus from the thetas in the weight function
 $F(\gamma,z;\lambda,m_i,m_j)$. Due to the proximity between the masses $m_2$ and $m_3$ given in Table~\ref{tableresidue} is not always apparent the several plateaus and superposition of them in the   $G_i$'s, which result in less structures than one could think.  
 On the other side, it is expected that the spectral densities will be smooth functions, and not having a discrete support, as in the model used here. In this case the plateaus would be smoothed out giving place, eventually, to oscillations, as a function of $\gamma$ and $z$. Another strict property of these weight functions, namely, the vanishing at the end-points due to $F(\gamma,\pm 1;\lambda,m_i,m_j)=0$ is of course shown in the figures~\ref{Fig:G1}-\ref{Fig:G4}, which leads  to the sharp ending peaks close to $z=\pm 1$ for some $\gamma$ values.

 \begin{figure}[thb]
\begin{center}
\vspace{0.5cm}
\epsfig{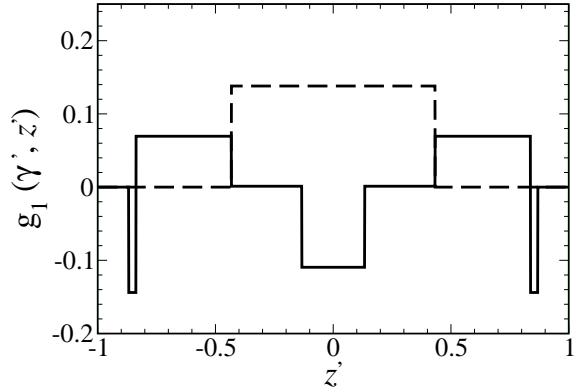} 
\caption{ Weight function $g_1(\gamma,z)$ as a function 
of $z$ for $\gamma=0.45$~GeV$^2$ coming from the 
covariant model with fixed constituent mass  (dashed line) 
and with the QCD pion inspired model (the single constituent mass pole model has $g_3=0$). For  the plot we use  the arbitrary value of $\mathcal{N}=100$, drop out the factor of $i$ and all parameters are in units of GeV.
} 
\label{Fig:g1}
\end{center} 
\end{figure}

\subsection{NIR: orthogonal basis}\label{subsec:NIRortho}

The Nakanishi weight functions of the  scalar amplitudes for the pion BS amplitude decomposed in the orthogonal basis are derived below using  Eqs.~\eqref{gg1}-\eqref{gg4}, and  having $G_i(\gamma,z)$ from Eqs.~\eqref{G1}-\eqref{G4}. The result is:
	\begin{equation}
	\begin{aligned}
	g_k(\gamma,z)=
&	\sum\limits_{1\leq i,j\leq 3}
\int_{0}^{\infty} d\gamma' \, {\widehat Y}^{(k)}_{ij}(\gamma,\gamma',z)\\	\times
&	\,F(\gamma',z;\lambda,m_i,m_j)\, ,
	\end{aligned}
		\label{ggk}
 	\end{equation}
where $k=$1, 2, 3 or 4. We have also introduced the functional
$ {\widehat Y}^{(k)}_{ij}(\gamma,\gamma',z)$ to simplify the notation. Note that at the end-points $g_k(\gamma,\pm 1)=0$,  as follows from $ F(\gamma,\pm 1;\lambda,m_i,m_j) =0$.

The functional associated with the weight function $g_1$ is given by:
\begin{multline}
 {\widehat Y}^{(1)}_{ij}(\gamma,\gamma',z) =w D_iD_j\,\theta(\gamma-\gamma')\,
(1-z\partial_z)
\\ +\Bigg[C_iC_j+D_iD_j\left(\frac{M^ 2_\pi}{4}-\gamma'\right)\Bigg]
\delta(\gamma-\gamma')\, ,
	\label{gg1m}
\end{multline}
which follows from Eqs.~\eqref{gg1},  \eqref{G1} and \eqref{G4}. It has the symmetry property,
\begin{equation}
 {\widehat Y}^{(1)}_{ij}(\gamma,\gamma',z)= {\widehat Y}^{(1)}_{ji}(\gamma,\gamma',z) \, ,   \end{equation}
which leads to 
$g_1(\gamma,-z)=g_1(\gamma,z)$, and we observe that the contribution from  $z\partial_z$ keeps the symmetry property under the transformation $z\to-z$ of the weight function. 

 \vspace{0.6cm}
 
 \begin{figure}[thb]
\begin{center}
\epsfig{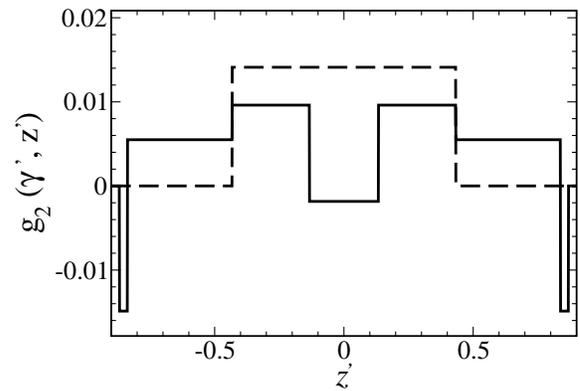} 
\caption{ Weight function $g_2(\gamma,z)$ as a function of $z$
for $\gamma=0.45$~GeV$^2$ coming from the covariant model 
with fixed constituent mass  (dashed line) and with the QCD pion inspired model (continuous line). For  the plot we use the arbitrary value of $\mathcal{N}=100$, drop out the factor of $i$ and all parameters are in units of GeV.} 
\label{Fig:g2}
\end{center} 
\end{figure}

In Fig.~\ref{Fig:g1}, we present the results for $g_1(\gamma,z)$ for $\gamma=0.45$~GeV$^2$ as a function of $z$, the characteristic plateaus are visible, as well as the vanishing at the end points. For comparison, we plot the weight function for the case where the quark propagator has only a  single pole at the constituent mass of $m_1=0.323$~GeV, while we kept the values of $\lambda$,  $D_1$ and $C_1$ in Table~\ref{tableresidue}. The single pole model presents just one plateau for $g_1$, a much simpler structure compared to the full model, which has three poles, with specific residua that makes vanish the weight function in the region where $g_1$, for the single pole model, is nonzero. The characteristic oscillations of the plateaus  of $g_1$ represents the interference between the residua with opposite signs and the overlapping  of support of the weight function $F(\gamma,z;\lambda,m_i,m_j)$, while the model with a constant constituent quark mass, has a much simple structure. 

\vspace{0.6cm}

 \begin{figure}[thb]
\begin{center}
\epsfig{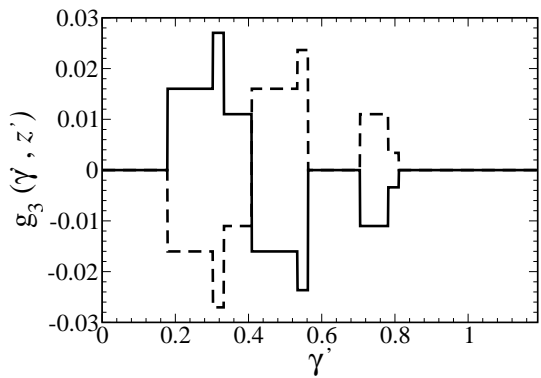} 
\caption{ Weight function $g_3(\gamma,z)$ as a function of $\gamma$ computed with the QCD pion inspired model for $z=0.5$ (continuous line) and for $z=-0.5$ (dashed line). For  the plot we use  the arbitrary value of $\mathcal{N}=100$, drop out the factor of $i$ and all parameters are in units of GeV.
}
\label{Fig:g3}
\end{center} 
\end{figure}

The functional that provides the Nakanishi weight function $g_2$ in Eq.~\eqref{ggk} is given by:
  \begin{multline}
{\widehat Y}^{(2)}_{ij}(\gamma,\gamma',z) =\frac{M_{\pi}}{2}~w\left[C_iD_j+C_jD_i\right] \delta(\gamma-\gamma')
\\ -\frac{2}{M_{\pi}}~w\left[D_iC_j-D_jC_i\right] 
\frac{\theta(\gamma-\gamma')}{M_\pi} ~ \partial_{z}z\, .
	\label{gg2m}
 	\end{multline}	
This functional is derived from Eqs.~\eqref{gg2} and \eqref{G2}, and it also leads to the symmetry property $g_2(\gamma,-z)=g_2(\gamma,z)$. In Fig.~\ref{Fig:g2}, we show the dependence on $z$ of $g_2(\gamma,z)$ for $\gamma=0.45$~GeV$^2$ considering the two models, the complete one with  three time-like  poles of the quark propagator, and the simplified one taking into account only the constituent mass pole corresponding to $m_1$. The general structure is similar to what we have observed for $g_1$ in Fig.~\ref{Fig:g1}, while the simplified model has only one plateau, as it is expected from the theta functions in $F(\gamma,z;\lambda,m_1,m_1)$, with the support for $m_1=0.323$~GeV being $0<\gamma \lesssim 1$~GeV$^2$.

The functional that builds the Nakanishi weight function $g_3$ is given by:
\begin{equation}
 {\widehat Y}^{(3)}_{ij}(\gamma,\gamma',z)=M_{\pi}w~[C_iD_j-C_jD_i]\,\delta(\gamma-\gamma')\, ,
 \label{gg3m}
 \end{equation}
 which follows from Eqs.~\eqref{gg3} and \eqref{G2}. It is proportional to $G_3-G_2$, which provides the odd property under the transformation $z\to-z$. Therefore, the plot of $g_3(\gamma,z)$ for $z=0.5$ as a function of $\gamma$ in Fig.~\ref{Fig:g3} complements the results shown for $G_3-G_2$ presented in the right panel of Fig.~\ref{Fig:G2G3}, and also provides one example of the $\gamma$ dependence in the Nakanishi weight function.  Fig.~\ref{Fig:g3} clearly reflects the scales involved in the model, namely the positions of the quark mass poles between 0.3 and 1~GeV (see Table~\ref{tableresidue}) as well as the pseudo-scalar vertex scale parameter $\lambda$ close to 1~GeV. Furthermore, the characteristic oscillations  observed 
 in the figure comes from the different signs of the residua, also as a consequence of the violation of the positivity constraints of the spectral densities of the scalar and vector parts of the quark propagator. Note that the single pole model results $g_3=0$. This indeed is a nice indication of the distinctive effect of the running quark mass model contrasting with the fixed constituent quark mass one.

 \begin{figure}[thb]
\begin{center}
\vspace{0.5cm}
\epsfig{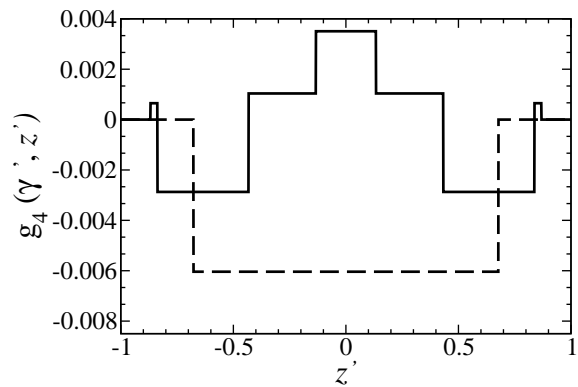} 
\caption{Weight function $g_4(\gamma,z)$ as a function of $z$
for $\gamma=0.45$~GeV$^2$ coming from the covariant model 
with fixed constituent mass (dashed line) and with the QCD pion inspired model (continuous line). For  the plot we use the arbitrary value of $\mathcal{N}=100$, drop out the factor of $i$ and all parameters are in units of GeV.
} 
\label{Fig:g4}
\end{center} 
\end{figure}
 
The functional associated with the Nakanishi weight function $g_4$ is derived through Eqs.~\eqref{gg4} and \eqref{G4} is given by: 
 \begin{equation}
 {\widehat Y}^{(4)}_{ij}(\gamma,\gamma',z)=~M_{\pi}^2\,w\,D_i\,D_j\,\delta(\gamma-\gamma')\, .
 \label{gg4m}
 \end{equation}
 The weight  function $g_4$ is proportional to $G_4$, which provides the even property under the transformation $z\to-z$. In this case, we plot in Fig.~\eqref{Fig:g4}, $g_4(\gamma,z)$ as a function of $z$ for $\gamma=0.45$~GeV$^2$ to compare with the result for the single quark constituent mass pole model. The single plateau of the latter is contrasted with the structure observed for the result obtained with the QCD inspired model.

\subsection{Normalization of the BS amplitude}\label{subsec:NBSA}

The normalization of the pion BS amplitude starts with  Eq.~\eqref{nap1b} contracted with $p_\mu$.  After performing the Dirac's algebra,  one gets that:
\begin{small}
\begin{multline}
iN_c\int\frac{d^4 k}{(2\pi)^4}\Big\{
\phi_1\phi_1+\phi_2\phi_2+
\mathcal{B}(k^2)\Big[\phi_3\phi_3+\phi_4\phi_4 \\ -4\phi_1\phi_4 \Big]
-\frac{2}{M_{\pi}}\Big[M(k_{\bar{q}}^2)+M(k_{q}^2)\Big]\phi_2\phi_1
\Big\}=1,
\label{integraltraco}
\end{multline}
\end{small}
where $\mathcal{B}(k^2)= \Big[(k\cdot p)^2 -k^2M_{\pi}^2\Big]/M_{\pi}^{4} $, $k_q=(k+p/2)$ and $k_{\bar q}=(k-p/2)$. 

The derivatives over the running masses  cancel out in the normalization due to the opposite signs of $p$ carried by the quark and antiquark  momentum in their arguments.
After  substituting the NIR of the scalar amplitudes $\phi_i$,   Eq.~\eqref{eq:NIRORTH}, in the normalization formula,  the analytic integration over  the loop momentum is easily performed using  standard Feynman parametrization. Finally, one finds that:
\begin{widetext}
\begin{equation}
    \begin{aligned}
 1 =&-~ \frac{3N_c}{ 32\pi^2}\int_{-1}^{+1} dz'\int_{0}^ {\infty} d\gamma' \int_{-1}^{+1} dz \int_{0}^{\infty} d\gamma
 \int ^1_0 dv 
 \left\{ ~v^2(1-v)^ 2~
 \left[ \frac{\mathcal{G}_{11}+\mathcal{G}_{22}  - 4\frac{m_0}{M_\pi} \mathcal{G}_{21}
 }{ \left[
 \frac{M_{\pi}^2}{ 4} \zeta'^2 +\gamma' v
+\gamma(1-v) \right]^4}
\right. \right. \\   & \left.\left. 
+\frac{\mathcal{G}_{33}+\mathcal{G}_{44}-
4~\mathcal{G}_{14}} { 2M_{\pi}^2\left[
\frac{M_{\pi}^2}{4} \zeta'^2 +\gamma' v
+\gamma(1-v)  \right]^3}\right]
-
\frac{16\,m^3}{M_\pi}\int_0^1du~
\frac{\mathcal{G}_{21}~(1-v-u)^2~u^2~\Theta(1-v-u)}{\left[\frac{M^2_\pi}{ 4}\zeta''^2-\frac{M^2_{\pi}}{4}+\lambda^2 +t\gamma+ u \gamma' \right]^5}
\right\},
\label{nend1}
\end{aligned}
\end{equation}
where  $\mathcal{G}_{ij} =g_i(\gamma ', z') \, g_j (\gamma, z) $, ~ $t=(1-v-u)$, 
~$
\zeta'=\big(vz'+(1-v)z\big)$ and 
$\zeta''=\big((1-v-u)z+uz'+v\big)\, .$
\end{widetext}

It is noteworthy to observe that the normalization  condition  of the BS amplitude, Eq.~\eqref{nend1}, contains all the Nakanishi weight functions, contrary to the valence wave function which does not have $g_1$.  This naive observation already suggests that the normalization condition contains contributions from components beyond the valence one. If $m^3=0$ the normalization reduces to the one found in~\cite{Paula_2021} for fixed constituent quark mass. The presence of the last term  is associated with the contribution of the running quark mass of this model to the BS normalization, curiously this term comes with the product of $\mathcal{G}_{21}$, which carries physics beyond the valence state, as $g_1$ is absent in this component of the wave function.

\section{ Valence probability and decay constant}\label{sec:valenceprob}
 
Once the BS amplitude is properly normalized,  we obtain the two spin components of   the valence  wave function, namely the spin antialigned one,      Eq.~\eqref{eq:psiantipar}, and the spin aligned one,
Eq.~\eqref{eq:psipar}. From the two spin components of the wave function, it is calculated the associated
valence probability $P_\text{val}$, Eq.~\eqref{eq:pval0}, the
corresponding spin decomposition 
$P_{\uparrow\downarrow}$
and $P_{\uparrow\uparrow}$, from Eqs.~\eqref{eq:Pupdown} and~\eqref{eq:Pupup}, respectively.
These quantities are not yet obtained for  the  QCD inspired model of the pion. The decay constant,  Eq.~\eqref{decayFL} are also computed, as a cross check of our formalism, giving that it was obtained in~\cite{Mello:2017mor} using a complete independent method.

 The calculations of these quantities are performed for different models, set I-IV, with and without  running quark mass.
 Set~I corresponds to the QCD inspired pion model, with the BS amplitude given by Eq.~\eqref{BSA1}. The sets II, III and IV  have fixed constituent quark mass corresponding to each of the poles of the dressed quark propagator, while keeping the same pseudoscalar vertex form. The normalization of the BS amplitude in these single pole models of the quark propagator is done with Eq.~\eqref{nend1}, disregarding the last term and substituting $m_0$ by the constituent quark mass.

\begin{table}[!thb]
 	\begin{center}
 \caption{Static pion properties ($m_\pi=140$MeV) for some sets of input parameters. The second column contains the constituent quark mass. The results for the following quantities are shown:  the valence probability $P_{val}$ (third column), and its decomposition in  the antialigned spins state, 
 $P_{\uparrow\downarrow}$ (fourth column), and aligned spins state, $P_{\uparrow\uparrow} $ (fifth column); the  pion decay constant (sixth column). }
 \vspace{0.2cm}
\begin{small}
\begin{tabular}{|c|c|c|c|c|c|}
\hline
Set  & Quark mass & $P_\text{val}$ & $P_{\uparrow\downarrow}$&
$P_{\uparrow\uparrow} $& $f_{\pi}$(MeV)     \\
\hline \hline
(I) & $M(k^2)$ [Eq.~\eqref{runningmass}] &0.70  &0.58  &0.12 &130.1  \\ 
\hline
(II)&  $m_1=323$ MeV     & 0.52  &0.41    &0.11  & 88
     \\ \hline
(III)& $m_2=645$ MeV     & 0.65  & 0.52   &0.13  &159
 \\ \hline
(IV)&  $m_3=954$ MeV     & 0.86  & 0.69   &0.17  &247 
 \\ \hline
Ref.~\cite{Paula_2021}&  $m= 255$ MeV   & 0.70  & 0.57   &0.13  &130 
 \\ \hline
\end{tabular}
\label{tabelainputs}
\end{small}
\end{center}
\end{table}

Our results are shown in Table~\ref{tabelainputs}. The set~I (pion model inspired by QCD)  reproduces the experimental value of the pion's decay constant,  $f_{\pi^{\pm}}^{exp .}=130.50(1)(3)(13)$~MeV~\cite{Zyla:2020zbs}, as well as the average value from LQCD calculations, $f_{\pi^{\pm}}^{LCQD }=130.2(1)(0.8)$~MeV~\cite{FLAG}. Furthermore, our result for $f_\pi$ agrees with  Ref.~\cite{Mello:2017mor}, where it was computed via integration of the LF momenta, which crosscheck the consistency of our treatment of the BS amplitude in terms of the NIR.  The valence probabilities in this model, as well as their decomposition in terms of the spin contributions were obtained.  The values found  are 0.7 for the total valence probability, 0.58 for the antialigned spin configuration and 0.12 for the aligned one.

We also present results for  the models with momentum-independent mass, corresponding to the three quark mass values of 323~MeV (set~II), 645~MeV (set~III) and 954~MeV (set~IV). The valence probability  $P_{val}$ and $f_\pi$ are in the range $[0.52 - 0.86]$ and $[98 - 247]$ MeV, respectively. This variation could be smaller if  $f_\pi$ fits the experimental value by  tuning the pseudoscalar vertex parameter.

It is interesting to compare with the results obtained by the solution of the BS equation in Minkowski space of Ref.~\cite{Paula_2021},
which includes one gluon exchange with mass of 637~MeV and a monopole quark-gluon vertex form factor with mass parameter of 306~MeV. From Table~\ref{tabelainputs}, one sees
that its results have  values for the valence probability and its spin decomposition  quite close to the pion model inspired by QCD. The valence probabilities, for
the anti-aligned and aligned spin, as well as in the dynamical model from~\cite{Paula_2021}, suggest the expected prevalence of the anti-aligned spins component of the pion wave function, a feature also exhibited in the single quark mass pole models.
Note that the contribution to the valence probability of  aligned spins component of the pion wave function is exclusively relativistic in nature, although it is by no means negligible, as the relative weight in the valence state amounts up to 24\%.
From the point of view of the expansion of the pion wave function on the  LF Fock space, we have that around 30\% of the normalization comes from higher Fock components, and therefore they are quite relevant in the description of the pion. Naively speaking, we should add that the remaining probability, $1 - P_{val}$, presumably is distributed between states of dressed degrees of freedom with few gluons and  quark-antiquark pairs beyond the valence state. 
However, we should convey that  it is still unclear how to build a basis in the LF Fock-space associated with the dressed quark quasi-particle.

On the other hand, for the BS amplitude models (II)-(IV)  we have that the constituent quark mass determines the binding energy ($B$), and larger values of $B = 2m_q - M_\pi$ corresponds to a more compact configuration for the pion. Therefore, the values of $f_\pi$ increases by increasing the quark mass, as we observe in Table~\ref{tabelainputs} an approximate proportionality between this two quantities (see also~\cite{Salcedo:2003yb}).
Furthermore, it is expected that the valence probability increases with the constituent mass, as the system tends to be non-relativistic and dominated by the valence configuration, which is clearly shown in the table II.

\section{Valence wave function}\label{Sec:LFprojampl}

\subsection{LF projected amplitudes}\label{subsec:LFPA}

We start by studying quantitatively 
the light-front amplitudes $\psi_i(\gamma,z)$,
Eq.~\eqref{wfi}, resulting from the projection of the scalar functions $\phi_i(k,p)$ from the orthogonal basis decomposition of the pion QCD inspired model BS amplitude. For convenience it is repeated below:
$$
 \psi_i(\gamma,z)=-\frac{\imath}{M_{\pi}}
 \int_{0}^{\infty}\hspace{-.2cm} d\gamma' ~
 \frac{g_i(\gamma',z)}{\left[\frac{z^2M_{\pi}^2}{4}+\gamma
 	+\gamma'\right]^2}\,.
$$
In the model, the weight functions $g_i$ have an upper-bound  in $\gamma$, which leads to the asymptotic behavior of $\psi_i(\gamma,z)$ by taking the limit of $\gamma=|\vec k_\perp|^ 2\to\infty$:
 \begin{equation}\label{psiasymp} \psi_i(\gamma,z)\sim -\frac{\imath}{\gamma^2}
\int_0^\infty \frac{d\gamma'}{M_\pi}g_i(\gamma',z)\, ,
 \end{equation}
 which for the QCD inspired model should mean in principle $\gamma\gg 1$~GeV$^2$ given the model parameters. However, as we are going to see, in practice it is enough that $\gamma\gg \Lambda_{QCD}^2$, with
  $\Lambda_{QCD}\sim 0.3$~GeV~\cite{Deur:2016tte},
 to have the asymptotic behavior. Indeed, such scale is not immediately obvious from the model parameters, it becomes apparent when we study the fall-off of the pion LF amplitudes and valence wave function with the transverse momentum. This point will be further explored when discussing the valence wave functions. 
 
 \begin{figure}[thb]
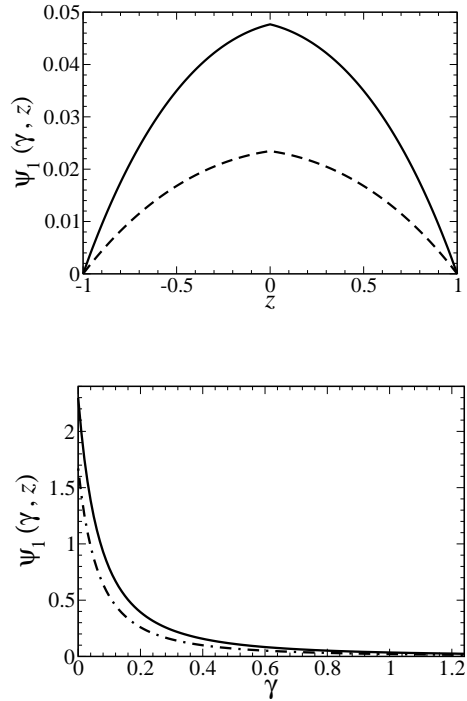

\begin{center}
\vspace{0.65cm}
\epsfig{figure=psi1novaf_z_.eps,width=6.0cm,angle=0} 
\vspace{0.35cm}
\vspace{0.60cm}
\epsfig{figure=psi1f_gamma_.eps,width=6.0cm,angle=0} 
\caption{Top panel: $\psi_1(\gamma,z)$ as a function of 
$z$ for $\gamma\,=\,1.0$~GeV$^2$ (solid line) and
1.5~GeV$^2$ (dashed line). 
Bottom panel: 
$\psi_1(\gamma,z)$ as a function of $\gamma$ [GeV$^2$] for $z\,=\, 0.5$ (solid line) and 0.75 (dash-dotted line).
 } 
\label{Fig:psi1}
\end{center} 
\end{figure}

\begin{figure}[thb]
\begin{center}

\vspace{0.5cm}

\epsfig{figure=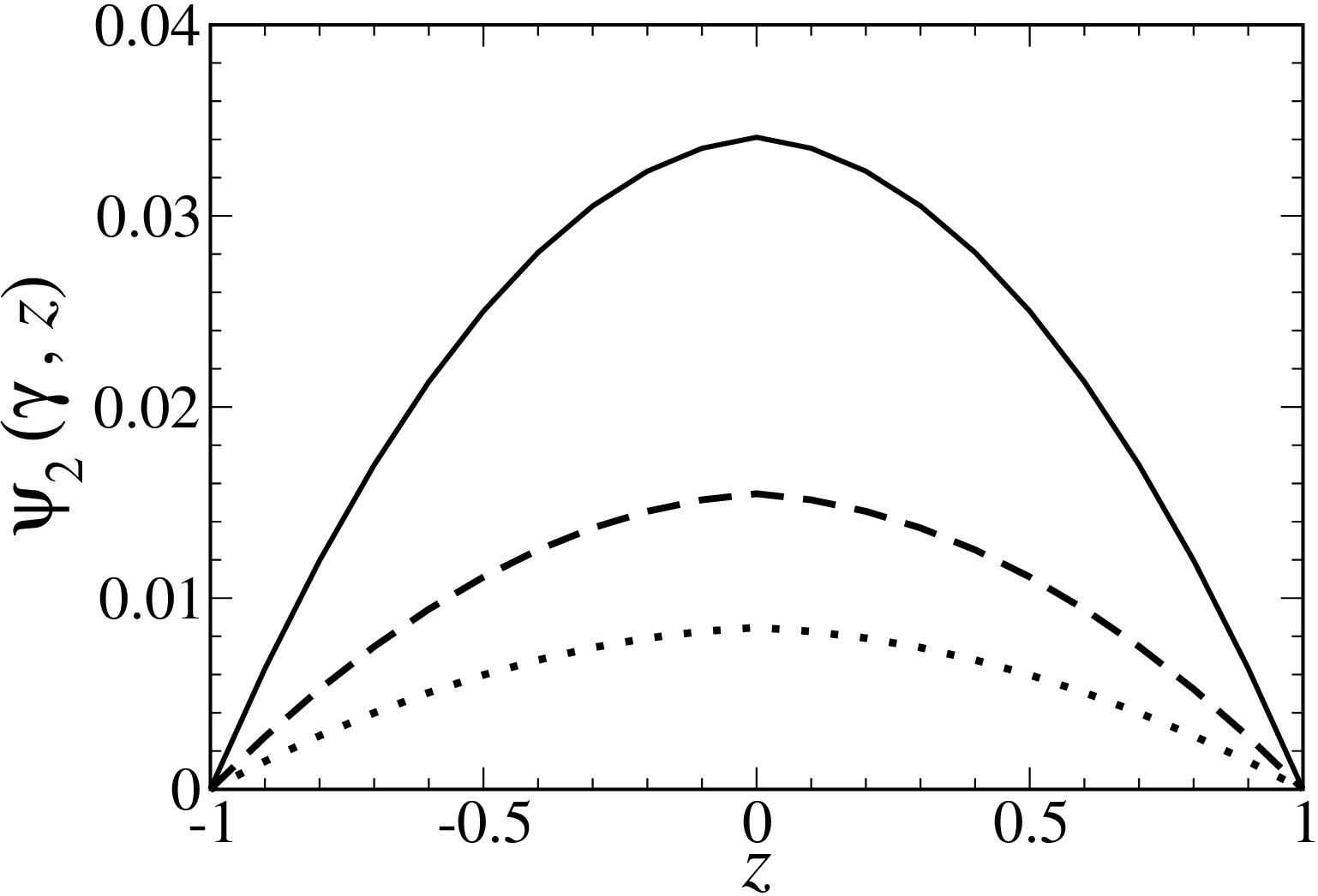,width=6.cm ,angle=0} 

\vspace{0.8cm}

\epsfig{figure=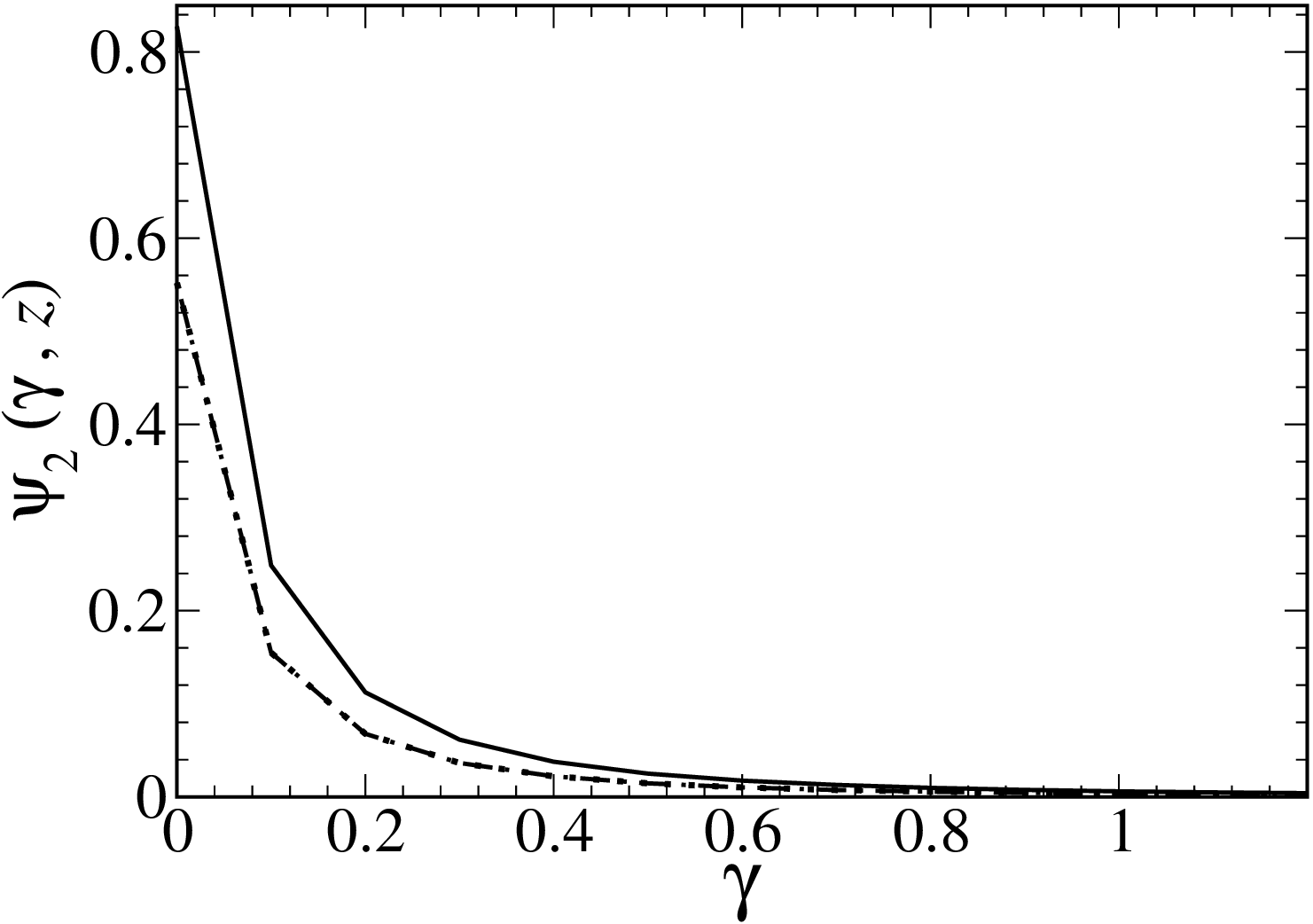,width=6.cm ,angle=0} 
\caption{ Top panel: $\psi_2(\gamma,z)$ as a function of $z$ for $\gamma\,=\,0.5$~GeV$^2$ (solid line), 0.75~GeV$^2$ (dashed line) and
1~GeV$^2$ (dotted line).
Bottom panel: $\psi_2(\gamma,z)$ as a function of $\gamma$ [GeV$^2$]
 for $z\,=\,0.5$ (solid line) and 0.75 (dash-dotted line).} 
\label{Fig:psi2}
\end{center} 
\end{figure}
 
\begin{figure}[thb]
\begin{center}

\vspace{0.5cm}

\epsfig{figure=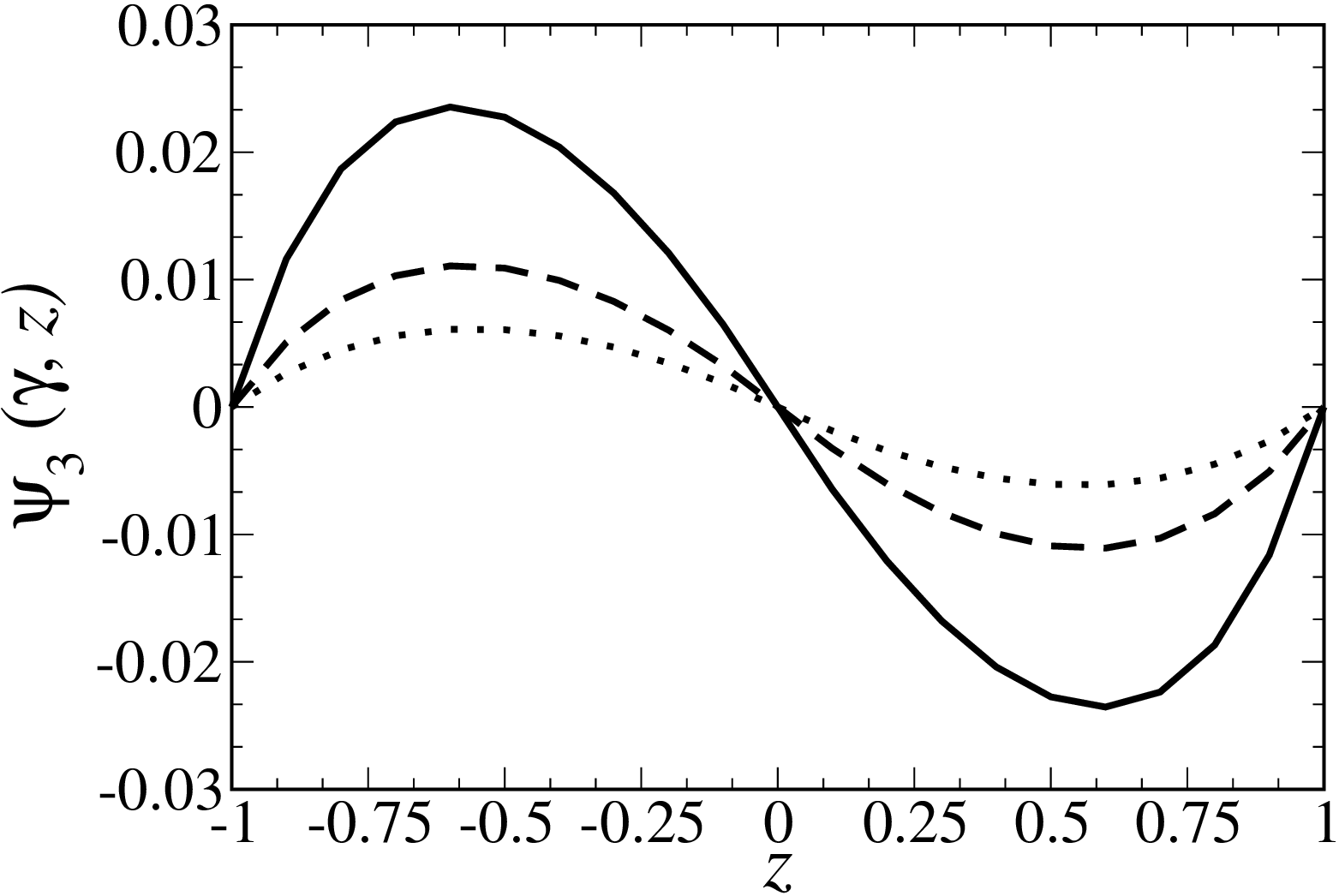,width=6.cm ,angle=0} 

\vspace{0.8cm}

\epsfig{figure=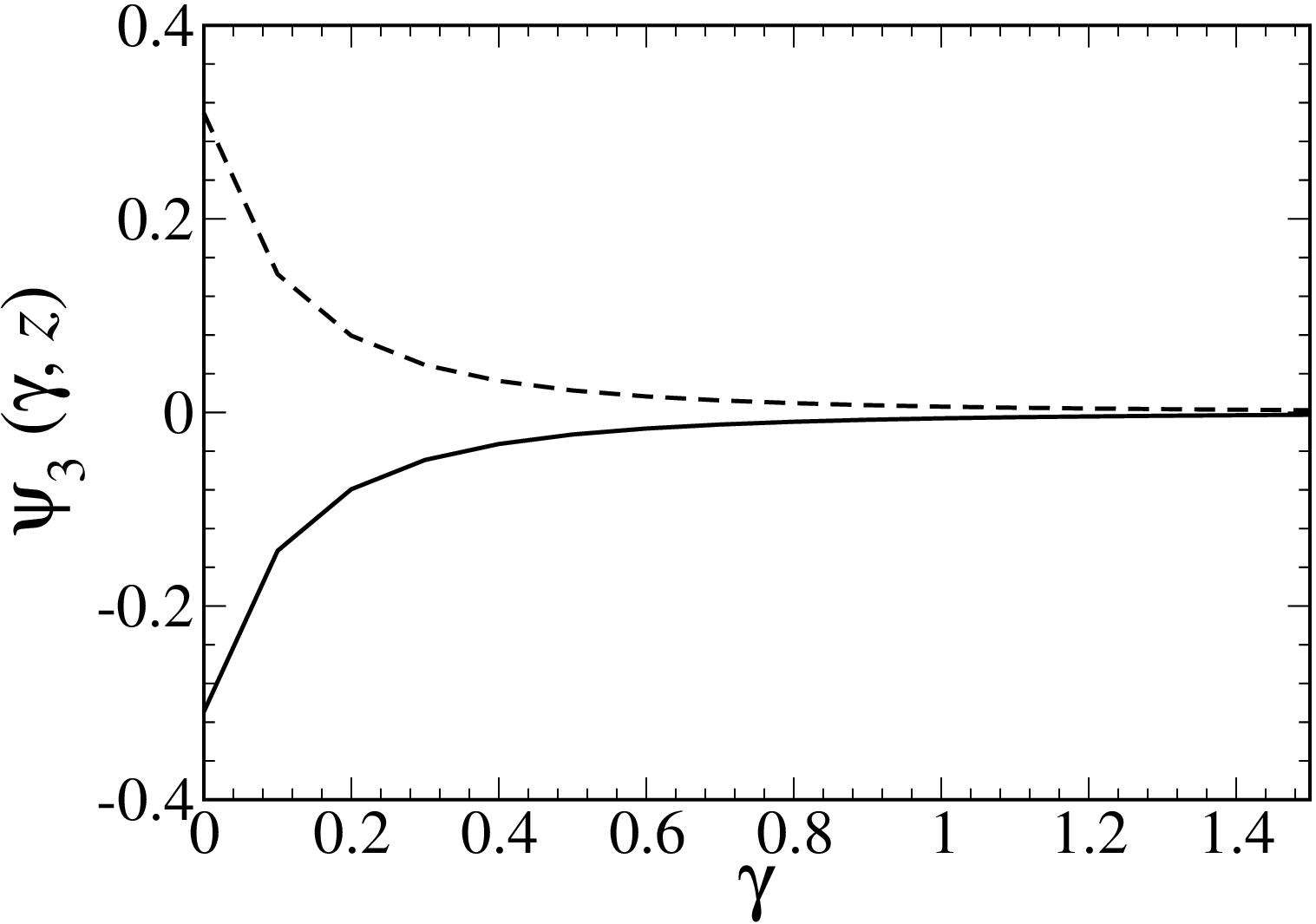,width=6.cm ,angle=0} 
\caption{Top panel:~$\psi_3(\gamma,z)$ as a function of $z$ for $\gamma =
0.5$~GeV$^2$ (solid  line), 0.75~GeV$^2$ (dashed line) and
1~GeV$^2$ (dotted line). Bottom panel: $\psi_3(\gamma,z)$ as a function of $\gamma$ [GeV$^2$]
for $z\,=\,0.5$ (solid line) and 0.75 (dash-dotted line).} 
\label{Fig:psi3}
\end{center} 
\end{figure}

\begin{figure}[thb]
\begin{center}

\vspace{0.5cm}

\epsfig{figure=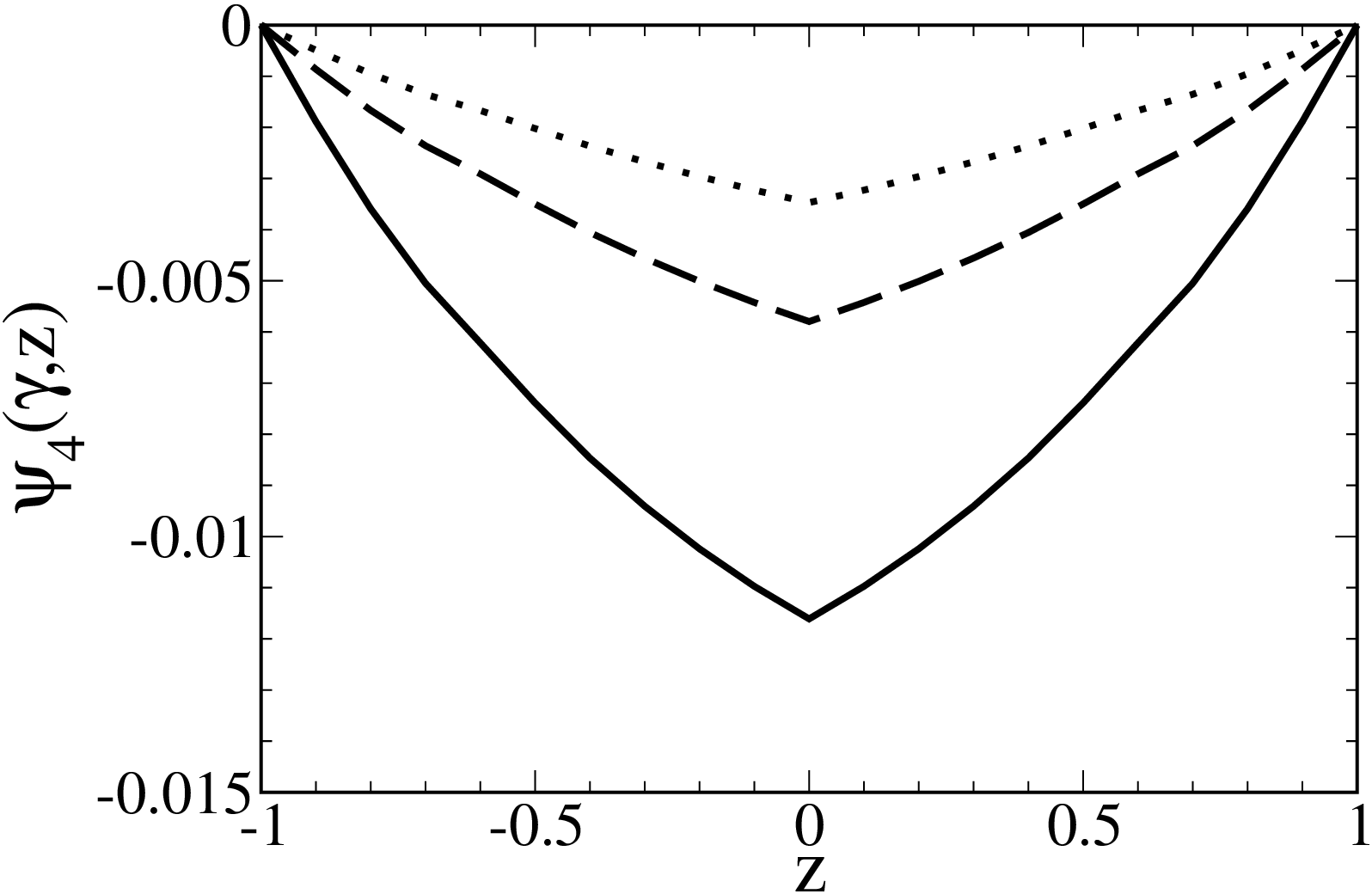,width=6.cm,angle=0} 

\vspace{0.8cm} 

\epsfig{figure=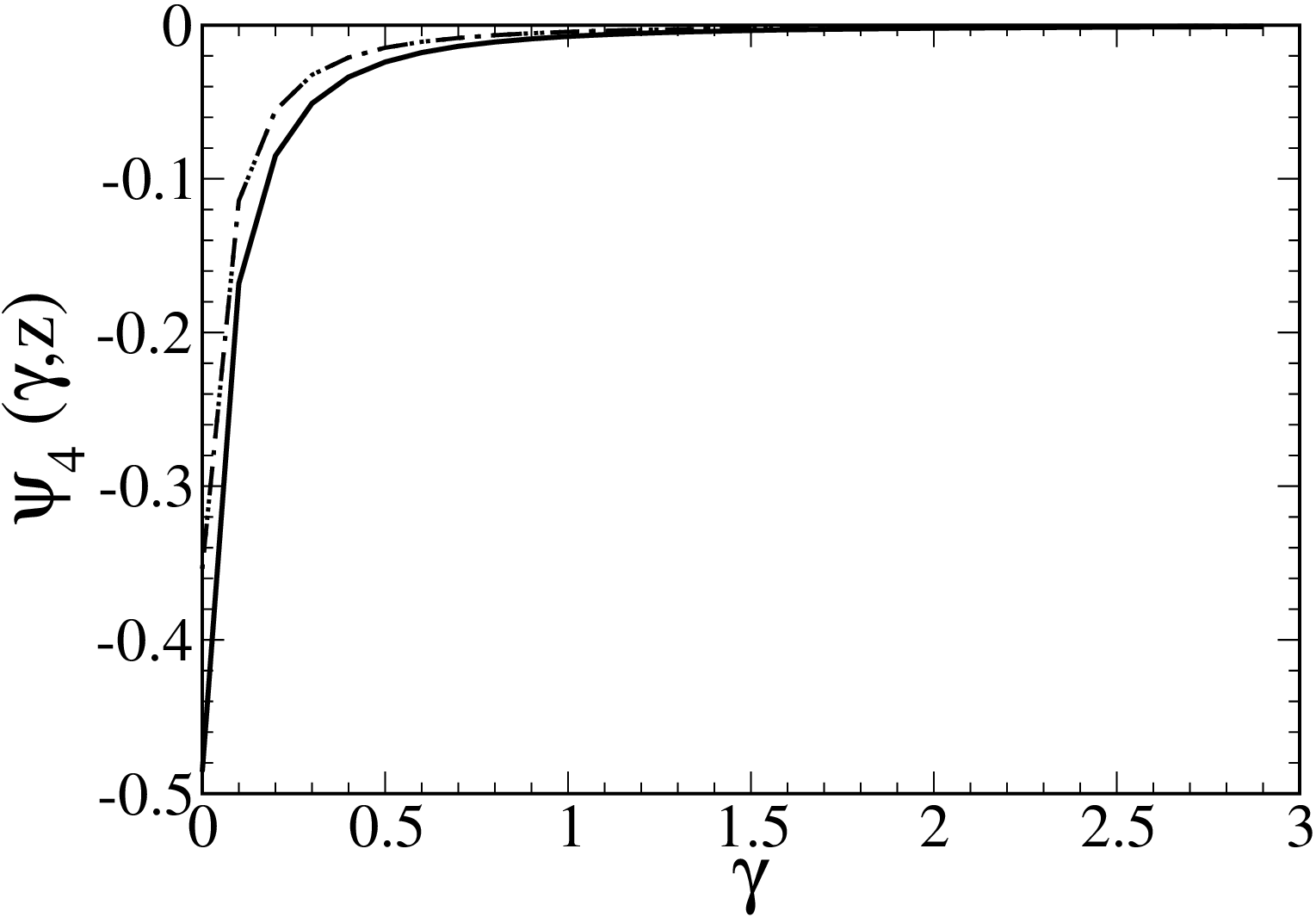,width=6.cm ,angle=0} 
\caption{Top panel: $\psi_4(\gamma,z)$ as a function of $z$
for $\gamma= 1$~GeV$^2$ (solid line), 1.5~GeV$^2$ (dashed line) and
2~GeV$^2$ (dotted line). Bottom panel: $\psi_2(\gamma,z)$ as a function of $\gamma$ [GeV$^2$] 
 for $z=0.5$ (solid line) and  0.75 (dash-dotted line). }
\label{Fig:psi4}
\end{center} 
\end{figure}

We present in Figs.~\ref{Fig:psi1} to \ref{Fig:psi4} the numerical results for the light-front project amplitudes  $\psi_i(\gamma,z)$ for the QCD inspired model.  The normalization of the weight functions are arbitrarily chosen as $\mathcal N$=1.
 
In Fig.~\ref{Fig:psi1} the dependencies of the amplitude $\psi_1(\gamma,z)$ in $z$ and $\gamma$ are shown in the top and bottom panels of the figure, respectively. The symmetry around $z=0$ reflects the charge conjugation symmetry of the pion state.  The behavior close to the end-points are shown  in the figure, and also the fast damping by increasing $\gamma$, as anticipated in Eq.~\eqref{psiasymp} with the asymptotic form $\sim \gamma^ {-2}$. Another clear feature is the scale of $\Lambda^2_{QCD}$ relevant for $\psi_1$ that drives the asymptotic behaviour with $\gamma\gg \Lambda^2_{QCD}$. Such asymptotic behavior is also verified in $\psi_j(\gamma,z)$ $(j=2,3,4)$ in figures~\ref{Fig:psi2}-\ref{Fig:psi4}. 

In Fig.~\ref{Fig:psi2}, the LF amplitude $\psi_2(\gamma,z)$ is plotted as a function of $z$ and $\gamma$  in the top and bottom panels of the figure, respectively. The symmetry property $z\to-z$ can be checked in the figure, as well as the behavior close to the end-points. 
 
In Fig.~\ref{Fig:psi3} the dependencies of the amplitude $\psi_3(\gamma,z)$ in $z$ and $\gamma$ are shown in the top and bottom panels of the figure, respectively.  The odd property of this amplitude with respect to the transformation $z\to-z$ is seen in the top panel of the figure, where results are presented for $\gamma=0.5$, 0.75 and 1~GeV$^2$. 

In Fig.~\ref{Fig:psi4}, the amplitude $\psi_4(\gamma,z)$ is shown and its dependencies in $z$ and $\gamma$ are illustrated in the top and bottom panels of the figure, respectively. The discontinuity of the derivative of $\partial_z\psi_4(\gamma,z)$  at $z=0$ is visible in the figure, while this feature of $\partial_z\psi_i(\gamma,z)|_{z=0}$  is smoothed out in $\psi_i(\gamma,z=0)$ with $i=(1,2,3)$. 
Such discontinuity is traced back to the discrete support of the  vector and scalar spectral densities of the quark propagator, which is expected to be washed out in a more realistic model.  
 
\subsection{Spin components}\label{subsec:spincomp}

The two spin components of the pion valence wave function, namely the anti-aligned and aligned ones are obtained from the LF projected amplitudes using Eqs.~\eqref{eq:psiantipar} and ~\eqref{eq:psipar}, respectively. These LF amplitudes for the QCD inspired pion model  were obtained in the previous section, where we made use of the LF projected scalar amplitudes $\phi_i(k,p)$ as written in Eq.~\eqref{wfi}.  In particular the LF projected amplitudes $\psi_i(\gamma,z)$ for $i=2$, 3 and 4, are the ones needed to compute the valence wave function. Note that the component $\psi_1$ does not directly contribute to these spin components, but  we remind that in the BS amplitude normalization condition~\eqref{nend1} all the scalar functions contribute.

\begin{figure}[thb!]
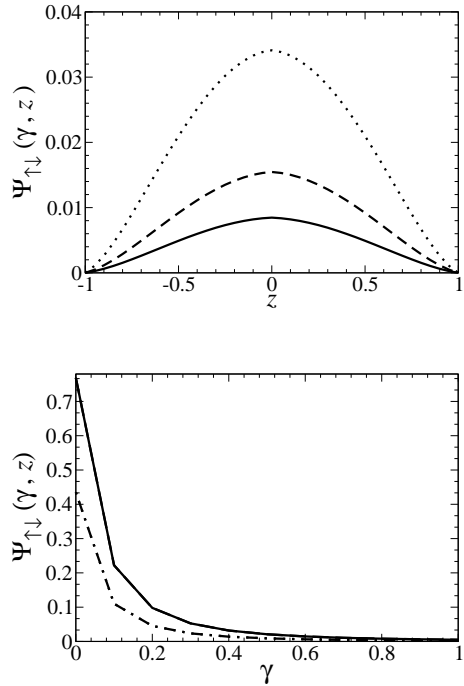

\begin{center}

\vspace{0.5cm}

\epsfig{figure=Psiantiparalelof_z_.eps,width=6.cm ,angle=0} 

\vspace{0.8cm} 

\epsfig{figure=Psiantiparalelof_gamma_.eps,width=6.cm ,angle=0} 
\caption{Top panel: antialigned quark spin component of the valence wave function,
$\Psi_{\uparrow\downarrow}(\gamma,z)$, as a function of $z$, for $\gamma=
0.5$~GeV$^2$ (dotted line), 0.75~GeV$^2$ (dashed line) and 1~GeV$^2$ (solid line).
Bottom panel: dependence on $\gamma$ [GeV$^2$] 
for $z=0.5$ (solid line) and 0.75  (dot-dashed line). 
Arbitrary normalization.} 
\label{Fig:psiantip}
\end{center} 
\end{figure}

\begin{figure}[thb!]
\begin{center}

\vspace{0.5cm}

\epsfig{figure=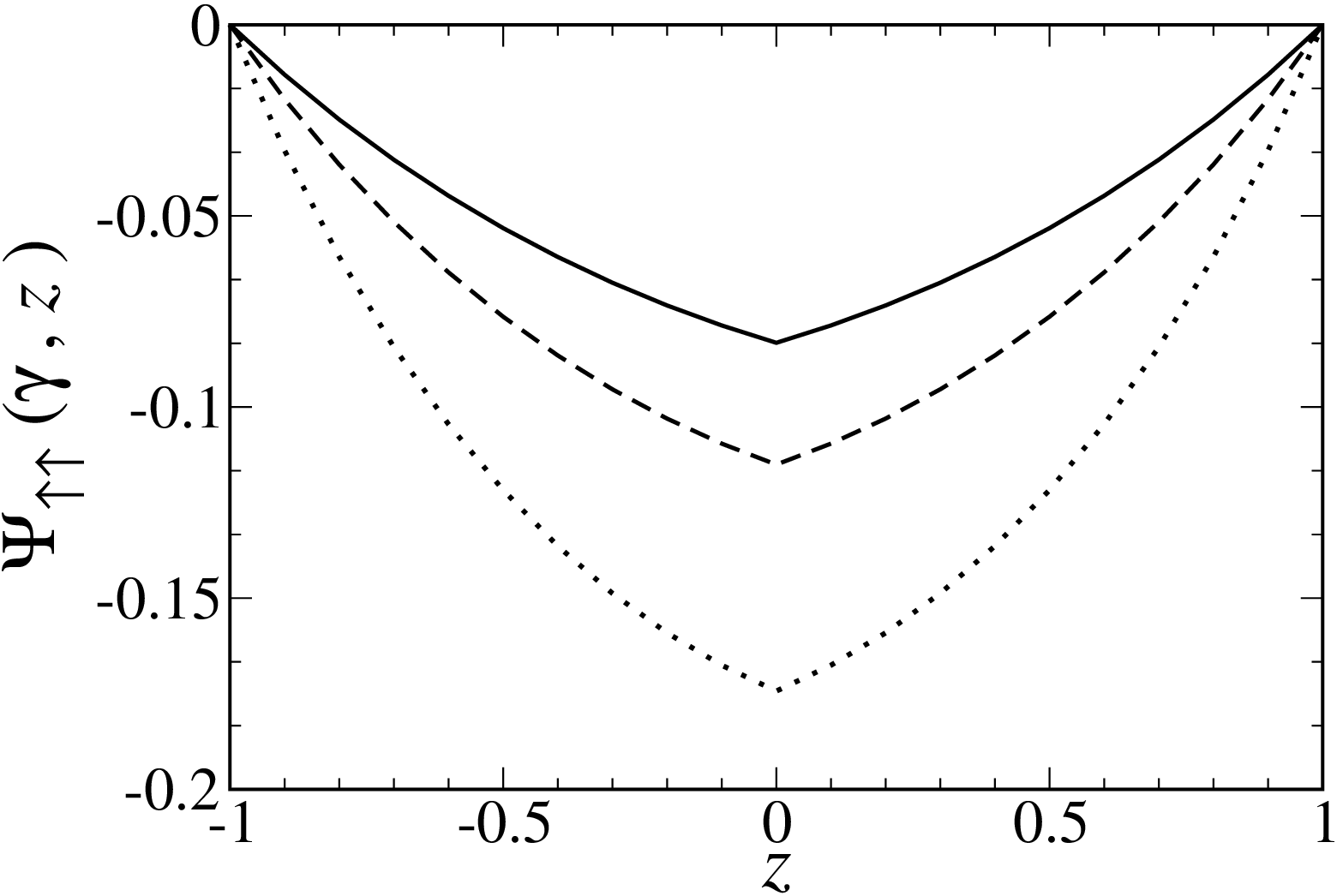,width=6.cm,angle=0} 

\vspace{0.8cm} 

\epsfig{figure=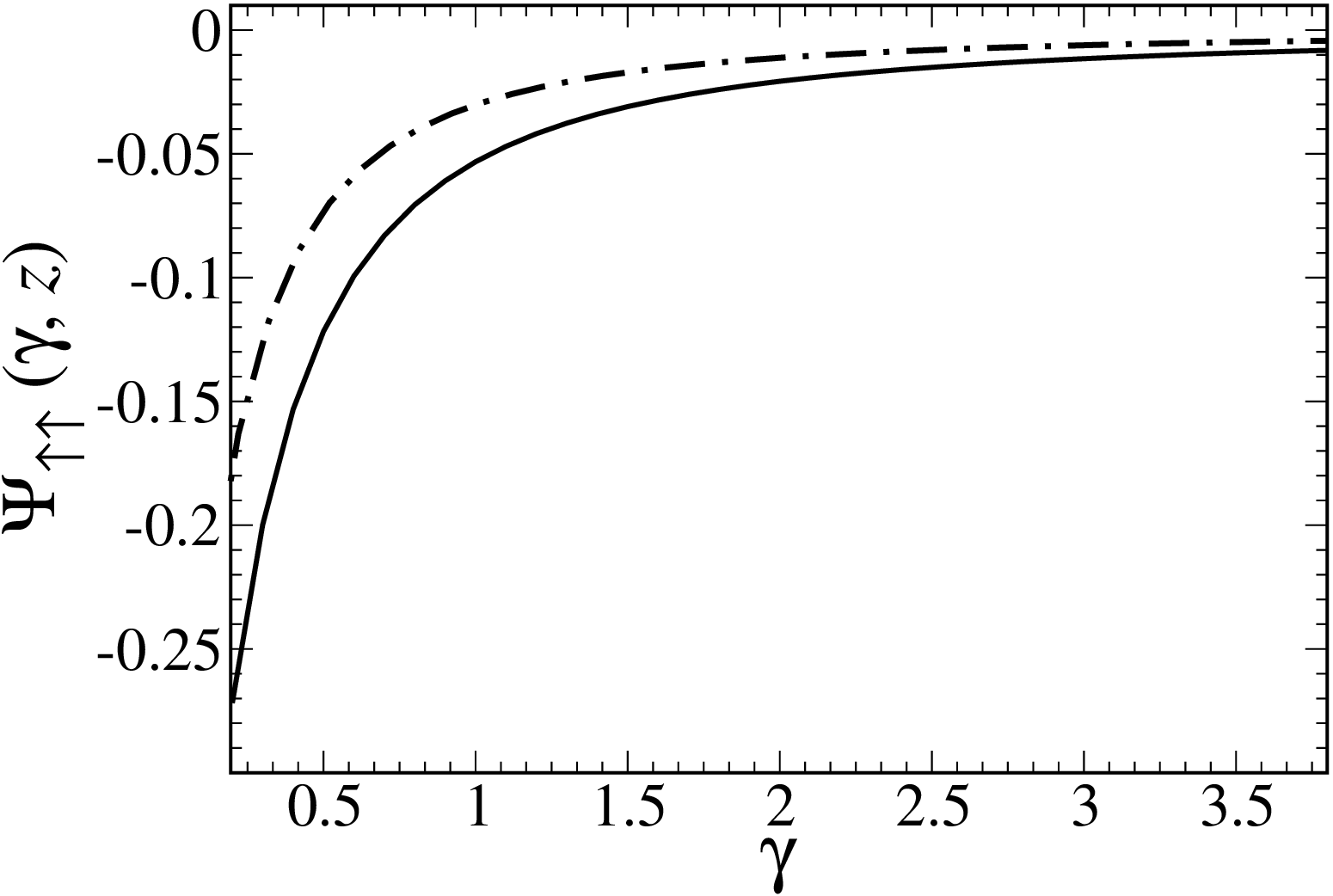,width=6.cm ,angle=0} 

\caption{Top panel: aligned quark spin component of the valence wave function,
$\Psi_{\uparrow\uparrow}(\gamma,z)$, as a function of $z$, for $\gamma=
0.5$~GeV$^2$ (dotted line), 0.75~GeV$^2$ (dashed line) and 1~GeV$^2$ (solid line).
Bottom panel: dependence on $\gamma$ [GeV$^2$] 
for $z=0.5$ (solid line) and 0.75  (dot-dashed line). 
Arbitrary normalization.} 
\label{Fig:psiparal}
\end{center} 
\end{figure}

In Figs.~\ref{Fig:psiantip} and \ref{Fig:psiparal}, we present the results for the valence wave function in the anti-aligned and aligned quark spin configurations, respectively.
In the top panel of both figures, the behavior with respect to $z$ near the end-points are visibly different  for the two components, where the anti-aligned spin one presents a faster falloff  towards $z\to\pm 1$ in  comparison to the aligned one. Associated with this property, we notice a narrower distribution of the former in relation to the latter.
These characteristics have been already  observed  for the pion wave function obtained within a dynamical model from the solution of the Minkowski space BS equation in the ladder approximation~\cite{Paula_2021}. The physical reason for the difference between the  end-point behavior is the relativistic origin of the aligned spin component.  Due to that  the quarks  in  this spin configuration have  to explore  the short-distance or UV region more frequently than when they are the in the anti-aligned spin state.  

The derivatives $\partial_z \psi_{\uparrow\downarrow}(\gamma,z)$ and $\partial_z \psi_{\uparrow\uparrow}(\gamma,z)$ are discontinuous at $z=0$ for the QCD inspired model, as seen in the top panels of Figs.~\ref{Fig:psiantip} and ~\ref{Fig:psiparal}, respectively. Such feature is smoothed for the antialigned spin component, while it is evident for the aligned spin component of the wave function. As we have already discussed in the previous subsection, such property is due to the theta functions appearing in Nakanishi weight functions. The discontinuity in the derivative is expect to disappear for a  spectral densities with a continuous support and not a  discrete one, as in the present model.

Another aspect found in the analysis of the LF amplitudes, $\psi_i$, and discussed in the previous subsection, is the emergent IR scale of $\Lambda_{QCD}^2$ in the $\gamma$ dependence of $\psi_{\uparrow\downarrow}$ and 
$\psi_{\uparrow\uparrow}$.
This IR scale corresponds to confinement distances of the order of 1~fm,  which indeed is corroborated by the value of 0.67~fm for the charge radius of the pion in this model obtained in Ref.~\cite{Mello:2017mor}.

 \begin{figure*}[thb]
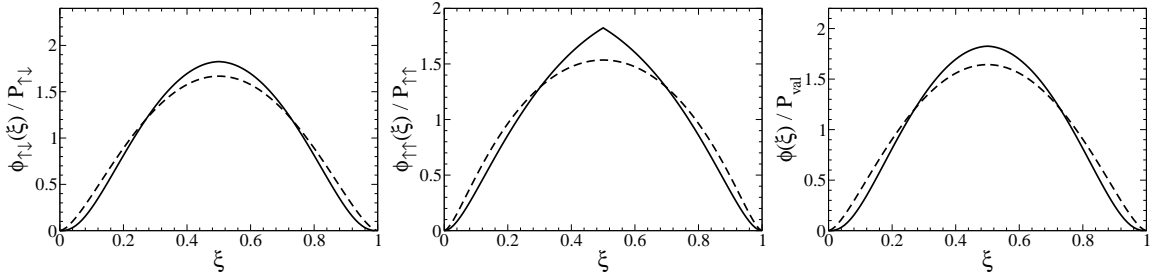

 	\begin{center}
\epsfig{figure=phiantiparxi.eps,width=5.cm,angle=0}
 \epsfig{figure=phiparalxi.eps,width=5.cm,angle=0} 
 \epsfig{figure=phifullxi.eps,width=5.cm,angle=0} 
 \caption{ Pion valence longitudinal-momentum distributions
 as a function of the momentum fraction for the QCD inspired model (solid line) and  for the solution of the BS equation in Minkowski space~\cite{Paula_2021}~(dashed line). Results presented for the antialigned spin configuration (left panel), aligned spin configuration (middle panel) and total (right panel).
}
 \label{Fig:spindecomp}
 \end{center} 
 \end{figure*}

The analytical form of the asymptotic behaviour of the spin components of the  wave function are  easily found by inspecting Eqs.~\eqref{eq:psiantipar} and ~\eqref{eq:psipar}, considering the finite support of $g_i(\gamma,z)$ in $\gamma$ and Eq.~\eqref{psiasymp}:
\begin{equation} \label{eq:PsiLFasymp}
    \begin{aligned}
   \Psi_{\uparrow\downarrow}(\gamma,z)&\sim
-\frac{\imath}{\gamma}\int_{0}^{\infty}\frac{d\gamma'}{M_{\pi}^3}
\frac{\partial}{\partial z}[g_3(\gamma',z)]\, ,
\\
\Psi_{\uparrow\uparrow}(\gamma,z)&\sim-\frac{\imath}{\gamma^\frac32} \int_0^\infty \frac{d\gamma'}{M_{\pi}}g_4(\gamma',z)\, .
    \end{aligned}
\end{equation}
The asymptotic behavior of the spin components of the valence LF wave function of the model, are consistent with the general form drived for QCD in Ref.~\cite{Ji:2003fw}, namely $~1/\gamma$ for the antialigned component  and $1/\gamma^\frac32$ for the aligned one. Note  in the last case the factor $\sqrt{\gamma}$ has been considered in the asymptotic behaviour in comparison with Ref.~\cite{Ji:2003fw}. By inspecting Figs.~\ref{Fig:psiantip} and \ref{Fig:psiparal} for $\gamma\gg\Lambda_{QCD}^2$ the asymptotic behavior of Eq.~\eqref{eq:PsiLFasymp} is confirmed. 
We should observe that the presence of the running mass was essential to make $g_3$ non vanishing. In the fixed constituent quark mass case, where $g_3=0$, the large transverse momentum fall-off of the antialigned spin component of the wave function would be $\sim 1/\gamma^2$ in disagreement with QCD.

 \begin{figure*}[thb]
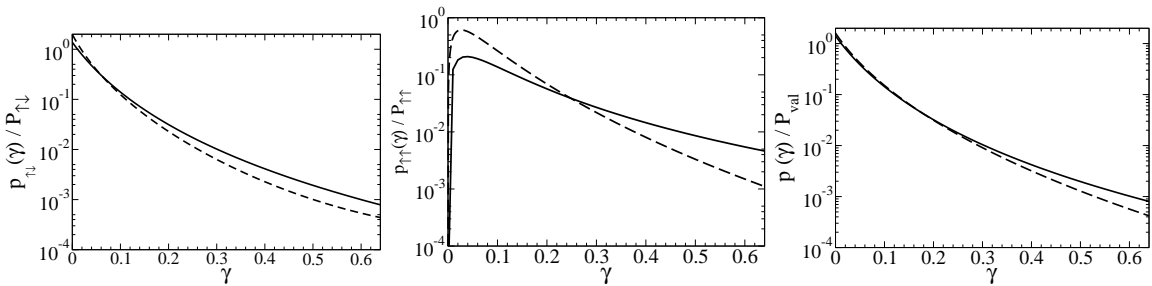

 	\begin{center}
\epsfig{figure=pantipargamma.eps,width=5.cm,angle=0} 
 \epsfig{figure=pparalgamma.eps,width=5.cm,angle=0} 
 \epsfig{figure=pfullgamma.eps ,width=5.cm,angle=0}  
 \caption{ Pion valence transverse-momentum distributions
 as a function of $\gamma=k^2_\perp$~[GeV$^2$] for the QCD inspired model (solid line) and  for the solution of the BS equation in Minkowski space~\cite{Paula_2021}~(dashed line). Results presented for the antialigned spin configuration (left panel), aligned spin configuration (middle panel) and total (right panel).}
 		\label{Fig:pantipargamma}
 	\end{center} 
 \end{figure*}

\subsection{Momentum distributions}\label{sec:result}

The longitudinal and transverse momentum distributions of the pion   valence LF wave function are found through the proper integration of the probability density, $\rho_{val}(\gamma,z)$ defined in Eq.~\eqref{eq:rhoval}, and its spin contributions $\rho_{\uparrow\downarrow}(\gamma,z)$ and $\rho_{\uparrow\uparrow}(\gamma,z)$ given in Eqs.~\eqref{eq:pvalud} and \eqref{eq:pvaluu}, respectively.

The  pion longitudinal momentum distribution decomposed in the individual contributions from  the two spin configurations, is written as
\begin{equation}
\phi(\xi)= \phi_{\uparrow\downarrow}(\xi)+\phi_{\uparrow\uparrow}(\xi)\, .
\label{somaphi}
\end{equation}
These quantities are the probability densities, 
integrated over the transverse momentum:
\begin{eqnarray} \label{eq:phiud}
&&\phi_{\uparrow\downarrow}(\xi)=\int^\infty_{0}d\gamma ~
\rho_{\uparrow\downarrow}(\gamma,z)\, , 
\\ \label{eq:phiuu}
&&\phi_{\uparrow\uparrow}(\xi)=\int^\infty_{0}d\gamma ~
\rho_{\uparrow\uparrow}(\gamma,z)\, ,
\end{eqnarray}
where we remind that $\xi=(1+z)/2$ is the quark longitudinal momentum fraction.

The results for the longitudinal momentum distributions are presented in Fig.~\ref{Fig:spindecomp}, which are  compared   to the results obtained with the solution of the ladder BS equation in Minkowski space that fits $f_\pi$ (model VIII in~\cite{Paula_2021}) and the experimental pion electromagnetic form factor~\cite{Ydrefors:2021dwa}.
For this comparison,  these distributions are normalized to 1, taking into account the respective probabilities given in Table~\ref{tabelainputs}. It should be noted that the distributions of the QCD-inspired model are narrower than those obtained with the BS solution from Ref.~\cite{Paula_2021}. This suggests the relevance of the dynamics of the BS equation in Minkowski space for that observable and on the  other side the quark dressing in the QCD  inspired model. In the last case at  the end-points   $\phi(\zeta)\sim\xi^\eta$ and $\sim(1-\xi)^\eta$, with $\eta\simeq 2$, while  $\eta=1.4$ for the solution of the BS equation in Ref.~\cite{Paula_2021}.

Another property observed for the longitudinal distribution of valence in Fig.~\ref{Fig:spindecomp} was the expected wider distribution for the aligned spin configuration  with respect to  the antialigned one. As we have already discussed such behavior comes as a  consequence of the genuinely relativistic nature of  the aligned spin component of the valence wave function. However, its  overall impact on the total valence distribution is small due to the low probability (see Table~\ref{tabelainputs}) associated with this  quark spin state.

The pion transverse momentum distribution is written in terms of  the   contributions from individual spin configurations as:
\begin{equation}
P(\gamma)= P_{\uparrow\downarrow}(\gamma)+P_{\uparrow\uparrow}(\gamma)\,,
\label{pgamma}
\end{equation}
where they are defined by integrating over $z$,   the spin antialigned and aligned momentum densities written in Eqs.~\eqref{eq:pvalud} and \eqref{eq:pvaluu}, respectively. 

In this way, the transverse momentum distributions, which complements our study of the valence state and its  spin decomposition, are given by:
\begin{eqnarray}\label{eq:Pud}
&&P_{\uparrow\downarrow}(\gamma)=\int^1_{-1} dz~
\rho_{\uparrow\downarrow}(\gamma,z)\, ,
 \\ \label{eq:Puu}
&&P_{\uparrow\uparrow}(\gamma)=\int^1_{-1} dz~\rho_{\downarrow\downarrow}(\gamma,z).
\end{eqnarray}

 We analyze in Fig.~\ref{Fig:pantipargamma}  the
distribution of the transverse momentum $P(\gamma)$ given in Eqs.~\eqref{pgamma}, and its decomposition into the spin configurations of the quarks written in Eq.~\eqref{eq:Pud} and  \eqref{eq:Puu}. For comparison purposes with the QCD-inspired model, we present in the figure the results obtained by solving the BS equation in Minkowski space in the ladder approximation~\cite{Paula_2021}. In general  the transverse momentum behavior of the two models is similar, with a falloff scale  of the order of $\Lambda_{QCD}$.  On overall, the aligned  spin transverse probability density in the region of $\gamma$  up  to 0.6~GeV$^2$ exhibits a slower falloff with respect to the  antialigned one, which reflects the sensitivity to the UV region, or short distances, this property is model independent. Independently of that, the contribution of the spin aligned state to the total is about 15\% of the total valence probability, as shown  in Table~\ref{tabelainputs}.

 \section{Conclusion}\label{sec:conclusion}
 
We summarize below the main results and discussions to settle down a theoretical framework for the Nakanishi integral representation of the pion BS amplitude within a chiral perspective, as well as  the application of the developed tools in practice to an example of a pion model inspired by QCD~\cite{Mello:2017mor}, where  its LF valence wave  function and momentum distributions were studied in detail. We expect that the developed framework can be an useful addition to further studies of the pion, where the self-energies are to be obtained dynamically from the solution of the Dyson-Schwinger equation in Minkowski space.

\begin{enumerate}[label=\roman*)]
\item 
A general framework was developed to build the Nakanishi integral representation of the Bethe-Salpeter amplitude of a model that incorporates the physics of spontaneous chiral symmetry breaking. 
 Technically, we take into account the K\"allen-Lehmann spectral decomposition of the quark propagator, and the integral representation of the pseudo-scalar vertex function. Starting from these basic ingredients then the Nakanishi integral representation~\cite{nak63} was build for the scalar amplitudes obtained in the decomposition of the BS amplitude in two basis of
 Dirac matrices, namely, one orthogonal, and the other one not.
 In particular, the first one was used in the solution of the BS equation for the pion in Minkowski space~\cite{Paula_2021}. The subtle transformation between the two sets of Nakanishi weight functions associated with the scalar amplitudes from the decomposition of the BS amplitude in the non-orthogonal and orthogonal basis was derived, which made possible our study. 
 
 \item We also introduced an auxiliary functional which given the spectral densities of the vector and scalar components of the quark propagator, and also the weight  function of the pseudo-scalar vertex, allows to derived the weight functions of associated to the BS amplitude. This formulation was restricted to the case where the density of the integral representation of the pseudo-scalar vertex function of the pion  was chosen, for simplicity, as having dependence only on a non-compact variable, which represents a scalar function of  the relative quark-antiquark momentum. This form is  suggested by the chiral limit where  the axial-vector Ward identities  imposes that the dominant component of the vertex function, the pseudo-scalar one, is  proportional to the quark scalar self-energy~\cite{CloPPNP14,horn2016pion}.
 
 \item 
 From the point of view of the  LF Fock-space decomposition of the pion wave function, the BS amplitude normalization receives contributions beyond the valence one taking into account an infinite sum of over the dressed quark-antiquark pairs with any number of dressed gluons, which have the pion quantum numbers. This is the counterpart of the well known LF quantization framework relying on the LF Fock-space basis built with eigenstates of the free LF Hamiltonian, where the normalization of the hadron  state is represented by the infinite sum over the probability of each Fock-state component of the wave function~\cite{BRODSKYPREP}.
 All this conceptual complexity that populates the public imagination, can be technically rephrased   by the projection onto the light-front of the Minkowski space Bethe-Salpeter equation~\cite{Sales:1999ec,Frederico:2010zh}. Although the  $\pi-q\bar q$ BS amplitude is defined in terms of the matrix element of only two quark operators, it contains the virtual propagation of the system in all LF Fock components with the pion  quantum numbers. On this picture, it remains the challenging task on how to build a LF Fock-space basis with quark and gluons dressed degrees of freedom. Despite of that, in our practical application  of the developed formalism to the pion state, we have made the initial step and studied its valence LF wave function with the dressed quark and antiquark degrees of freedom using the integral representation method developed within the described framework.

\item In the practical application of the tools developed  in  the present formal framework, we have study the valence LF wave function properties in detail within the pion model  inspired by QCD. In this model the Bethe-Salpeter amplitude incorporates  the quark self-energy, where the mass function is momentum dependent, and this in turn determines the quark-pion pseudoscalar vertex at the chiral limit (zero quark current mass), assuming the theoretical validity of the chiral Ward-Takahashi identities. The model has a  satisfactory phenomenology reproducing the pion  decay constant and electromagnetic form factor~\cite{Mello:2017mor}.
The quark propagator has three time-like poles  at 323,  645 and 954~MeV corresponding to spectral densities of the  vector and scalar components of the propagator with a discrete support, which also happens for the form factor of the  pseudo-scalar $\pi-q\bar q$ vertex that has a time-like  pole at 846~MeV. Here,  we have derived the analytic form of the Nakanishi weight functions, which has as an essential ingredient the auxiliary functional built in the development of our general framework. We found that many of the properties of the weight functions are due to the particular form of the auxiliary  functional, like the vanishing at the end-points, and a rich structure, due  to the alternate signs of the residue associate with the  spectral densities  at the time-like poles of the   dressed quark propagator model. Choosing  a fixed constituent quark mass, such rich  structure is missed in the Nakanishi weight  functions.

\item 
We computed from the integral representation of the Bethe-Salpeter amplitude the pion light-front valence wave function decomposed in terms of quark helicities coupled in antialigned and aligned spin states.  As a cross-check of the developed formalism, after the normalization of the Bethe-Salpeter amplitude, we computed  the pion decay constant,  and it reproduced the original model results obtained  in~\cite{Mello:2017mor}, with a completely independent method.
We have computed  the  probabilities of the  quark spin aligned state  and antialigned  spin one in the valence wave function. We found quite unexpected that the valence probability of 0.70, divided in 0.58  for the  antialigned spin state and 0.12 for the aligned one.
Curiously this result agrees with the probabilities found from the solution of the pion Minkowski space Bethe-Salpeter equation in the ladder approximation, where  the quark-gluon vertex scale parameter, the quark and gluon masses have values suggested by lattice QCD calculations and  tuned to fit the decay constant.

 \item  
 We studied the structure of the pion valence LF wave function, by means of its spin decomposition, namely their  dependence on the transverse and longitudinal  momentum fraction. The  typical scale for the falloff of the wave functions is found to be  of the order of $\Lambda_{QCD}$ although not immediately visible from the parameters of the model. Furthermore, we found that the two spin components  fulfill the asymptotic power-law falloff as $1/k_\perp^2$ and $1/k_\perp^3$,  for the spin antialigned and aligned wave  functions, respectively, in agreement  with the counting rule for hadronic light-cone wave functions derived for QCD~\cite{Ji:2003fw}.
 
\item  The dressing of the quarks kept the property that the valence longitudinal momentum distribution in the aligned spin state is wider than the antialigned one,  due to  the relativistic origin of the former. Such property was also found from the solution of the BS amplitude in Minkowski space~\cite{Paula_2021}. The behavior at the  end-points, namely $\xi^\eta$  and $(1-\xi)^\eta$, for $\xi\to 0$  and $\xi\to 1$, respectively, has the exponent $\eta\simeq 2$ at the pion scale. In addition, the  valence transverse momentum distribution was studied.
 
\end{enumerate}
 
{\it Acknowledgements:}~
TF is grateful to Giovanni Salm\`e for helpful discussions on the Haag's theorem implications.
This work was partially supported by Coordena\c c\~ao de Aperfei\c coamento de Pessoal de N\'ivel Superior  (CAPES) under the grant 88881.309870/2018-01 (WP),  Funda\c c\~ao de Amparo \`a Pesquisa 
do Estado de S\~ao Paulo (FAPESP)  grants 2017/05660-0 and 2019/07767-1 (TF),  2019/02923-5 (JPBCM), and by Conselho Nacional de Desenvolvimento Cient\'\i fico 
e Tecnol\'ogico (CNPq) grants 308486/2015-3 (TF),  307131/2020-3 (JPBCM), 313030/2021-9 (WP), and 464898/2014-5 (INCT-FNA).

\vspace{3.0cm}

\newpage

\appendix
\section{Elimination of $k^2$ for $g_1(\gamma',z')$} 
\label{appendixa}
The Lorentz invariant~$k^2$ in the numerator of Eq.~(\ref{confusao}) for $\phi_1(k,p)$ has to be manipulated before applying the principle of uniqueness of the Nakanishi integral representation in order to extract $g_1(\gamma',z')$. For this aim, we use the following identity:
\begin{equation}
\label{eq1:appa}
k^2=(k^2+z\,k\cdot p-\gamma)-z\,k\cdot p+\gamma \, .
\end{equation}

This allows to write:
\begin{eqnarray}\label{eq2:appa}
&&\int_{-1}^1dz\int_{0}^{\infty} d\gamma
\frac{k^2\,G_4(\gamma,z)}{\left[k^2+z\,k\cdot p-\gamma + \imath \epsilon\right]^3} = \nonumber \\
&&\int_{-1}^1dz\int_{0}^{\infty} d\gamma\frac{G_4(\gamma,z)}{\left[k^2+z\,k\cdot p-\gamma + \imath \epsilon\right]^2}
\nonumber \\
&&-(k\cdot p)\int_{-1}^1dz\int_{0}^{\infty}
d\gamma\frac{z\,G_4(\gamma,z)}{\left[k^2+z\,
k\cdot p-\gamma + \imath \epsilon\right]^3} \nonumber\\
&&+\int_{-1}^1dz\int_{0}^{\infty}d\gamma \frac{\gamma\,G_4(\gamma,z)}{\left[k^2+z\,k\cdot p - \gamma + \imath \epsilon\right]^3} \, .
\end{eqnarray}
We integrate by parts in $\gamma$ the first term of the right-hand side of Eq.~\eqref{eq2:appa} in order to write the NIR form with power 3 in the denominator, as given below:
\begin{equation}\label{eq3:appa}
2\int^ 1_{-1} dz\int_0^{\infty}d\gamma
\frac{\int_{0}^{\gamma}d\gamma'\,G_4(\gamma',z)}{(k^2+z\,k\cdot p-\gamma + \imath \epsilon)^3}\, .
\end{equation}
For the second term in the right-hand side of Eq.~\eqref{eq2:appa}, we integrate by
parts in $z$, which results in: 
\begin{equation}\label{eq4:appa}
-\frac12\int_{0}^{\infty}d\gamma\int_{-1}^1 dz
\frac{\partial_{z}[z\,G_4(\gamma,z)]}
{[k^2+z\,k\cdot p-\gamma + \imath \epsilon]^2}\, ,
\end{equation}
where  the boundary term is set to zero. Then, in order to have power 3 in the denominator of Eq.~\eqref{eq4:appa} we use the same integration by parts as we have done to derive Eq.~\eqref{eq3:appa}:
\begin{equation}\label{eq5:appa}
-\int_{0}^{\infty}d\gamma\int_{-1}^1 dz
	\frac{\int_{0}^\gamma d\gamma'\, \partial_{z}[z\,G_4(\gamma',z)]}
	{[k^2+z\,k\cdot p-\gamma + \imath \epsilon]^3}\, .
	\end{equation}
Finally, using Eqs.~\eqref{eq2:appa}-\eqref{eq5:appa} one arrives at Eq.~\eqref{gg1}.

\end{document}